\documentclass[manuscript]{aastex}

\usepackage{lscape,longtable,booktabs}
\usepackage{multirow}

\newcommand\sw{{\it Swift}}
\newcommand\fe{{\it Fermi}}
\newcommand\TN{{$T_{\,90}$}}
\newcommand\Epeak{{$E_{\rm peak}$}}
\newcommand\Eiso{{$E_{\rm iso}$}}

\def\go{\mathrel{\raise.3ex\hbox{$>$}\mkern-14mu\lower0.6ex\hbox{$\sim$}}}
\def\lo{\mathrel{\raise.3ex\hbox{$<$}\mkern-14mu\lower0.6ex\hbox{$\sim$}}}
\def\simeq{\mathrel{\raise.3ex\hbox{$\sim$}\mkern-14mu\lower0.4ex\hbox{$-$}}}
\def\deg{{${}^\circ$}}
\shorttitle{Broadband spectroscopy of high--energy GRBs}
\shortauthors{Bissaldi et al.}

\begin{document}


\title{First--year Results of Broadband Spectroscopy \\
       of the Brightest Fermi--GBM Gamma--Ray Bursts}


\author{Elisabetta Bissaldi\altaffilmark{1,2},
    Andreas~von Kienlin\altaffilmark{1},
    Chryssa~Kouveliotou\altaffilmark{3},
    Michael~S.~Briggs\altaffilmark{4},
    Valerie~Connaughton\altaffilmark{4},
    Jochen~Greiner\altaffilmark{1},
    David~Gruber\altaffilmark{1},
    Giselher Lichti\altaffilmark{1}, \\
    P.~N.~Bhat\altaffilmark{4},
    J.~Michael~Burgess\altaffilmark{4},
    Vandiver~Chaplin\altaffilmark{4},
    Roland~Diehl\altaffilmark{1},
    Gerald~J.~Fishman\altaffilmark{3},
    Gerard~Fitzpatrick\altaffilmark{5},
    Suzanne~Foley\altaffilmark{1},
    Melissa~Gibby\altaffilmark{6},
    Misty~Giles\altaffilmark{6},
    Adam~Goldstein\altaffilmark{4},
    Sylvain~Guiriec\altaffilmark{4},
    Alexander~J.~van~der~Horst\altaffilmark{7},
    Marc~Kippen\altaffilmark{8},
    Lin~Lin\altaffilmark{4}, \\
    Sheila~McBreen\altaffilmark{5},
    Charles~A.~Meegan\altaffilmark{7},
    William~S.~Paciesas\altaffilmark{4},
    Robert~D.~Preece\altaffilmark{4},
    Arne~Rau\altaffilmark{1},
    Dave~Tierney\altaffilmark{5}
    and Colleen~Wilson--Hodge\altaffilmark{3}}
\altaffiltext{1}{Max-Planck-Institut f\"ur extraterrestrische Physik, Giessenbachstrasse 1, \\ 85748 Garching, Germany}
\altaffiltext{2}{{\it Current address:} Institute of Astro and Particle Physics, University Innsbruck, \\ Technikerstrasse 25, 6176 Innsbruck, Austria}
\altaffiltext{3}{Space Science Office, VP62, NASA/Marshall Space Flight Center, 
Huntsville, \\ AL 35812, USA}
\altaffiltext{4}{University of Alabama in Huntsville, 
320 Sparkman Drive, Huntsville, AL 35809, USA}
\altaffiltext{5}{University College, Dublin, Belfield, 
Stillorgan Road, Dublin 4, Ireland}
\altaffiltext{6}{Jacobs Technology, Inc., Huntsville, Alabama}
\altaffiltext{7}{Universities Space Research Association, 
320 Sparkman Drive, Huntsville, \\ AL 35809, USA}
\altaffiltext{8}{Los Alamos National Laboratory, 
PO Box 1663, Los Alamos, NM 87545, USA}
%
%

%
\begin{abstract}
We present our results of the temporal and spectral analysis 
of a sample of 52 bright and hard gamma--ray bursts (GRBs) observed 
with the Fermi Gamma--ray Burst Monitor
(GBM) during its first year of operation ( July 2008 -- July 2009). Our sample
was selected from a total of 253 GBM GRBs based on the event peak count 
rate measured between 0.2 and 40\,MeV. The final sample comprised 34 long 
and 18 short GRBs. These numbers show that the GBM sample contains a much 
larger fraction of short GRBs, than the {\it CGRO}/BATSE data set, which we explain 
as the result of  our (different) selection criteria and the improved GBM trigger 
algorithms, which favor collection of short, bright GRBs over BATSE. 
A first by--product of our selection methodology is the determination 
of a detection threshold from the GBM data alone, above which GRBs 
most likely will be detected in the MeV/GeV range with the Large 
Area Telescope (LAT) onboard \fe. This predictor will be very useful 
for future multiwavelength GRB follow ups with ground and space based 
observatories. Further we have estimated the burst durations up to 10\,MeV 
and for the first time expanded the 
duration--energy relationship in the GRB light curves to high energies. 
We confirm that GRB durations decline with energy as a power law with 
index approximately $-0.4$, as was found earlier with the BATSE data 
and we also notice evidence of a possible cutoff or break at 
higher energies. Finally, we performed time--integrated spectral 
analysis of all 52 bursts and compared their spectral parameters 
with those obtained with the larger data sample of the BATSE data. 
We find that the two parameter data sets are similar and confirm 
that short GRBs are in general harder than longer ones. 
\end{abstract}
\section{Introduction}
The most exciting results in the last decade of gamma--ray 
burst (GRB) science at MeV--GeV energies came from the 
combined observations of the Burst And Transient Source Experiment \citep[BATSE,][]{FIS93}
with the Energetic Gamma--Ray Experiment Telescope \citep[EGRET,][]{FIC94}. 
Both instruments operated between 1991
and 2000 on board the Compton Gamma--Ray Observatory ({\it CGRO}) and
covered the energy bands from 20\,keV to 20\,MeV and from 
20\,MeV to 30\,GeV, respectively. 
BATSE observed 2704 bursts \citep{PAC99}, providing
the largest GRB database from a single experiment
thus far. Out of this sample, only five bursts were
detected with EGRET above 100\,MeV and only
one of these, GRB~930131 \citep{SOM94}, had high--energy
emission that was consistent with an extrapolation
from its spectrum obtained with BATSE between 
25~keV and 4~MeV \citep[see also][]{DIN03}. 
However, later analysis of the combined data from the 
BATSE/Large Area Detectors (LADs) and the EGRET calorimeter, 
the Total Absorption Shower Counter (TASC), 
uncovered an MeV component in GRB~941017 described by 
a power law with photon index approximately $-1$ up 
to about 200\,MeV \citep{GON03}. \citet{GON09} subsequently 
searched the TASC data for a response to 68 bright BATSE bursts. 
They found that only 21 showed emission detectable by 
TASC and of these, only three contained spectra with 
peak energy, \Epeak~$>$~2\,MeV. As these spectra were only 
found in a time--resolved analysis, \citet{GON09} 
claimed that the high energy component could be hidden in the 
brightness of the synchrotron emission in a time--integrated spectrum. 
They suggested that the existence of the high 
energy component indicated additional non--thermal processes at the source. 

With the successful launch on 2008 June 11 of 
the \fe\ Gamma--ray Space Telescope ({\it FGST}, hereafter \fe), it is now possible 
to search and confirm the spectral signatures 
from GRBs up to very high energies. \fe\ is an international 
and multi--agency space observatory that studies the Cosmos 
over an unprecedented broad energy range (10\,keV to 300\,GeV).
The Gamma--Ray Burst Monitor (GBM) is the secondary instrument onboard {\it Fermi},
operating between $\sim$8~keV and $\sim$40~MeV. The search for
higher energy GRB emission is carried out with the primary instrument, the
Large Area Telescope \citep[LAT, ][]{ATW09}, a pair conversion telescope, like EGRET, operating 
in the energy range between 20\,MeV and 300\,GeV.
GBM alone provides a much wider energy coverage than any other
current GRB mission, such as \sw\ \citep{GEH04}. As of November 2010, 
joint high--energy GBM/LAT observations 
have been carried out for 19 GRBs above 100\,MeV \citep[see e.g.][]{PEL11}.  This number represents 
$\sim$3\%\ of the total number of GRBs observed with GBM.

The small number of LAT GRB detections may be due to instrumental bias. 
The combined GBM/LAT GRB spectra are usually well--described in the MeV--GeV
range by a single power law with an index in the approximate range of $-1$ to $-3$
\citep[see e.g.][]{ABD09a}.
This result is in agreement with the distributions of the high--energy power--law indices observed
with BATSE in the $\sim$30~keV--$\sim$2~MeV energy
range \citep[][hereafter K06]{KAN06}. Consequently, photon counts above $\sim$1~MeV
are usually very low, and this, combined with the limited 
Field--of--View (FoV) of the LAT, results
in much fewer GRBs observed in the multi--MeV band than
in the keV--band. Currently, detailed calculations of 
LAT upper limits are being performed for the brightest 
bursts detected with GBM \citep{ABD11a}. 

In this paper we establish a well--defined sample of 
52 bright GBM GRBs with broad--band
spectral coverage and with statistically significant 
high spectral resolution.  All events were collected
during the first year of GBM operations (July, 14, 2008 -- July 15, 2009) 
and were detected up to MeV
energies with GBM alone. In Section \ref{Sec_Instrum} we discuss the GBM 
instrumentation and data types and in Section \ref{Sec_Samp_Selec} we describe 
the selection methodology for our sample, mainly based on the 
peak count rates measured above 500 keV.  In Section \ref{Sec_Temp} we 
present the temporal analysis of our sample
over several energy bands. In Section \ref{Sec_Spec} we describe the 
results of our time--integrated spectral analysis using 
different photon models to fit each spectrum, and discuss the 
distributions and correlations of the spectral parameters 
for the best models. Our conclusions are presented in Section \ref{Sec_Concl}.
\section{Instrumentation and Data Types}\label{Sec_Instrum}
The primary role of GBM was to augment the science
return from \fe\ in the study of GRBs by making observations
at lower energies ($\sim$8\,keV to $\sim$40\,MeV) and thus bridging the gap with those of the LAT.
The GBM flight hardware comprises a set of 12 Thallium--doped
Sodium Iodide crystals (NaI(Tl), hereafter NaI) and 
two Bismuth Germanate crystals
(Bi$_4$Ge$_3$O$_{13}$, commonly abbreviated as BGO).
The individual NaI detectors are mounted around the spacecraft in four groups of three. Their 
arrangement results in an exposure of the whole sky unocculted by the Earth in orbit.
The NaI detectors are able to detect $\gamma$--rays in the energy 
range between $\sim$8~keV and $\sim$1~MeV. 
The two BGO detectors are mounted on opposite sides of the \fe\ spacecraft.
With their energy range between $\sim$0.2 and $\sim$40~MeV, they
provide the overlap in energy with the LAT instrument
and are crucial in the study of high--energy, hard bursts.

To trigger the GBM flight software (FSW), two or more NaI detectors
must have a simultaneous statistically significant rate increase 
above the background rate (usually $>5\sigma$).
This requirement increases the
threshold against statistical fluctuations and suppresses
triggering due to non--astrophysical events that appear in only one detector, such as
phosphorescence spikes.
Before performing any spectral analysis, the detector geometry 
with respect to the GRB direction must be carefully taken
into account. Detectors which see the burst at an angle $>$50\deg,
or which suffer from blockages (by the solar panels, by the LAT or by the spacecraft
itself) were excluded. Sometimes these criteria result in a single NaI detector
to be chosen for the spectral analysis,
which is then fitted together with the mostly illuminated BGO detector.
The best detector combination for each burst is given in columns 4 (NaIs) and 5
(BGO) of Table~\ref{Tab_Prop_All}.

All three GBM data types, namely CSPEC, CTIME and TTE data, 
were used for the analysis presented in the following sections.
A detailed description of these data types can be found
in \citet{MEE09}. 
CTIME and CSPEC data were used in burst--mode, i.~e.~at 64~ms and 1.024~s temporal 
resolution, respectively.
The BGO peak count--rate analysis and the determination of the burst durations
in the integrated BGO energy range
are based on CTIME files, which have the finest temporal resolution (64~ms)
with modest energy resolution consisting of 8 energy channels. For the 
determination of the duration dependence on energy, 
we used TTE data both for the NaI and the BGO detectors. 
CSPEC and TTE data provide an energy resolution consisting of
128 energy channels and were used for all spectral analyses. 
The coarse time--resolution CSPEC data are normally used as pre--trigger background data
for the TTE data, since the latter only include $\sim$~30~s 
before the trigger time. TTE data are then collected up to 300~s post--trigger, and
in all cases discussed hereafter cover the whole burst duration.
For each GRB, the data type used for spectral analysis
is listed in column 7 of Table~\ref{Tab_Prop_All}.
\section{Sample Selection Methodology}\label{Sec_Samp_Selec}
We selected our sample based on two requirements: (i) a significant
count rate excess above background ($>$~3$\sigma$) 
measured by the most illuminated NaI detectors in
the 50--300 keV energy range, to ensure good statistics; and (ii)
a significant count rate excess above background  
measured by the most illuminated BGO detector in the 500~keV--1~MeV
during the main burst emission episode (T$_{90}$).
This combination allows a broadband spectral analysis
of GBM data, which spans about three decades in energy (10~keV--10~MeV).
It also represents a significant difference from previous selection
criteria adopted by K06, which were mainly based on
peak photon flux and fluence values measured in the 50--300 keV energy range.
Criterion (ii) was afforded by the good performance of the BGO
detectors at those energies ($>$~400~keV), where the effective
area of the NaI detectors rapidly decreases \citep{BIS09}. Below, we 
discuss the methodology employed for the burst sample selection.

The first coarser burst selection was based on the analysis of the
GBM telemetry packets, which are automatically produced 
during a trigger and contain all trigger information
such as locations, classifications and accumulated rates \citep{MEE09}.
The so--called GBM ``maximum rates'' 
observed over a short period after trigger time ($<$4~s) 
are produced from the accumulations made for the trigger algorithms
and are evaluated as statistical significance (signal over noise ratio; SNR) 
versus the background.
Typically, the FSW background interval ranges from 
about $-36$~s to $-4$~s with respect to the trigger time, thus
excluding the most recent few seconds of data and avoiding in most cases the
contamination by pre--trigger data from the burst.
We automatically selected bursts showing an increase of more than 80 counts/s 
over background in at least one BGO detector over the full BGO energy
range ($\sim$250\,keV to ~$\sim$40\,MeV).

The refined burst selection was based on the analysis
of BGO CTIME light curves. As previously
mentioned, CTIME data have a 64~ms temporal resolution
during burst--mode and spectral resolution of 
8 energy channels.
Channel edges are controlled using the specific Lookup Tables (LUTs), 
which map the 4096 raw channels into the 8 energy channels 
\citep{MEE09}. Exact channel boundaries can vary from detector
to detector (BGO 0 or 1) and from burst to burst.
The BGO CTIME background was computed including pre-- and post--trigger
time intervals, usually from $-300$~s to $+300$~s in case of long bursts 
and from $-50$~s to $+50$~s for short bursts,
excluding the interval of the burst itself.
The background--subtracted light curve was then examined for 
the maximum or ``peak'' count rate on the 64~ms--timescale
over each individual CTIME energy channel. 

The resulting total number of GRBs included in this
spectral analysis is 52, which approximately represents
$\sim$20\% of all bursts detected during the first year of GBM operation.
These bursts are listed in Table~\ref{Tab_Prop_All}.
The first three columns list the GBM trigger number, the GRB name\footnote{The naming
convention follows the
Gamma--ray bursts Coordinates Network (GCN) publication policy. Bursts which were 
not reported in a GCN circular are not given a name in column 2.}, and the burst
trigger time (in MET). The numbers of the NaI (from 0 to 11) and BGO (0 and 1)
detectors used for the temporal and spectral analysis are reported
in columns 4 and 5. Column 6 gives the angle ($\theta$) of the burst
with respect to the LAT boresight. The LAT FoV covers those
events which are located at $\theta<$~65\deg. This value
represents the initial angle from the source calculated at trigger time
and can vary during the burst in the case 
of a slew of the spacecraft (the so--called autonomous repoint recommendation 
or ARR). The GBM data type and time interval (with respect to
the burst trigger time, $T_0$) adopted for the spectral
analysis are listed in the last three columns.

The full sample of 52 bursts was further subdivided
according to the detection significance of the event peak in a BGO energy channel.
GRBs detected with more than 3~$\sigma$ significance in the first six
BGO energy channels constitute the corresponding {\it Channel} sample.
All bursts in the full sample are detected in 
{\it Ch.0} ($\sim$200--500 keV) and {\it Ch.1} ($\sim$500--1000 keV),
28 bursts are detected in  {\it Ch.2} ($\sim$1--2 MeV), 
14 bursts are detected in  {\it Ch.3} ($\sim$2--5 MeV), and
6 bursts are detected in {\it Ch.4} ($\sim$5--10 MeV).

\begin{figure}
\centering
\includegraphics[width=0.5\textwidth,bb=0 0 540 720,clip]{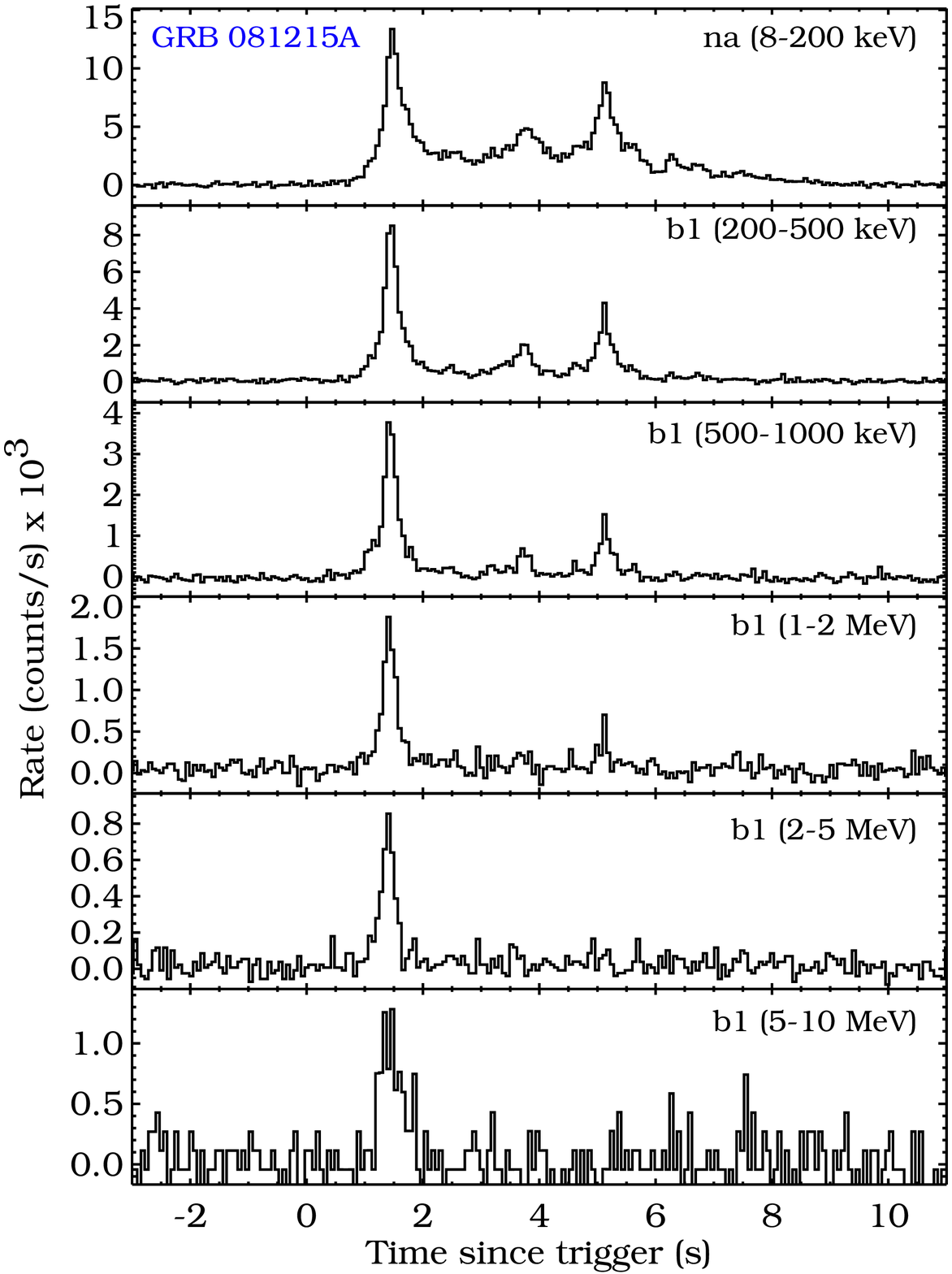}
\caption{Background--subtracted light curves of the long GRB~081215
observed with the GBM detectors. 
The panels show the sum of the 
counts in different energy bands as seen
by ({\it top histogram}) the most illuminated NaI detector in 
the~8--200~keV energy range, and  ({\it bottom five histograms}), 
the BGO detectors covering the first five CTIME energy channels. 
The bin width is 128~ms.}
\label{LC_081215}
\end{figure}

Figure~\ref{LC_081215} shows an example light curve of one of the brightest 
GBM bursts in our sample, namely the long GRB~081215A,
which is detected up to $>$ 5 MeV in BGO.
The {\it top panel} shows the 8--200 keV band, covered by the most 
illuminated NaI detector(s). The other panels show the BGO light 
curve in different energy ranges, covering 
five CTIME energy channels ({\it Ch.0--Ch.4}). 
This very bright burst was also marginally
detected by the LAT.
Indeed, $\theta \sim$86\deg,  which 
means that neither directional nor 
energy information could be obtained 
with the standard analysis procedures.
However, \citet{PEL10} recently presented a new technique
to recover the signal from the GRB prompt emission between $\sim$30\,MeV and 
100\,MeV, which differs from the standard LAT analysis
(the so--called ``LAT Low--Energy'' technique, or LLE).
Using such non--standard data selection, 
over 100 events above background 
are detected within a 0.5\,s interval 
in coincidence with the main GBM 
peak. The significance of this excess 
is greater than 8$\sigma$ \citep[see also][]{MCE08}.
\subsection{BGO Effective--Area Correction}
In order to correct for the dependence of the BGO effective area
on the incidence angle, we calculated an additional 
scaling factor accounting for the angle between the burst position
and the BGO detectors. This was mainly necessary since the peak count rate
analysis was performed on data without taking the instrument response into account.
In \citet{BIS09}, the off--axis response of the BGO detectors was 
measured at different energies for both flight module detectors at various angles between
0\deg\ (i.e., on axis) and 90\deg. A scaling factor could be calculated
for each incident direction and then used to correct
the peak count rates. The correction factors are relatively small
out to $\sim$40\deg\ and strongly increase toward 90\deg.
At higher energies, the correction factor is not as high as
at lower energies. This mainly reflects the strong absorption
of low--energy photons by the BGO photomultiplier tubes (PMTs).
\subsection{LAT Detections}
\begin{figure}
\centering
\begin{tabular}{c}
\includegraphics[width=0.5\textwidth,bb=0 0 590 482,clip]{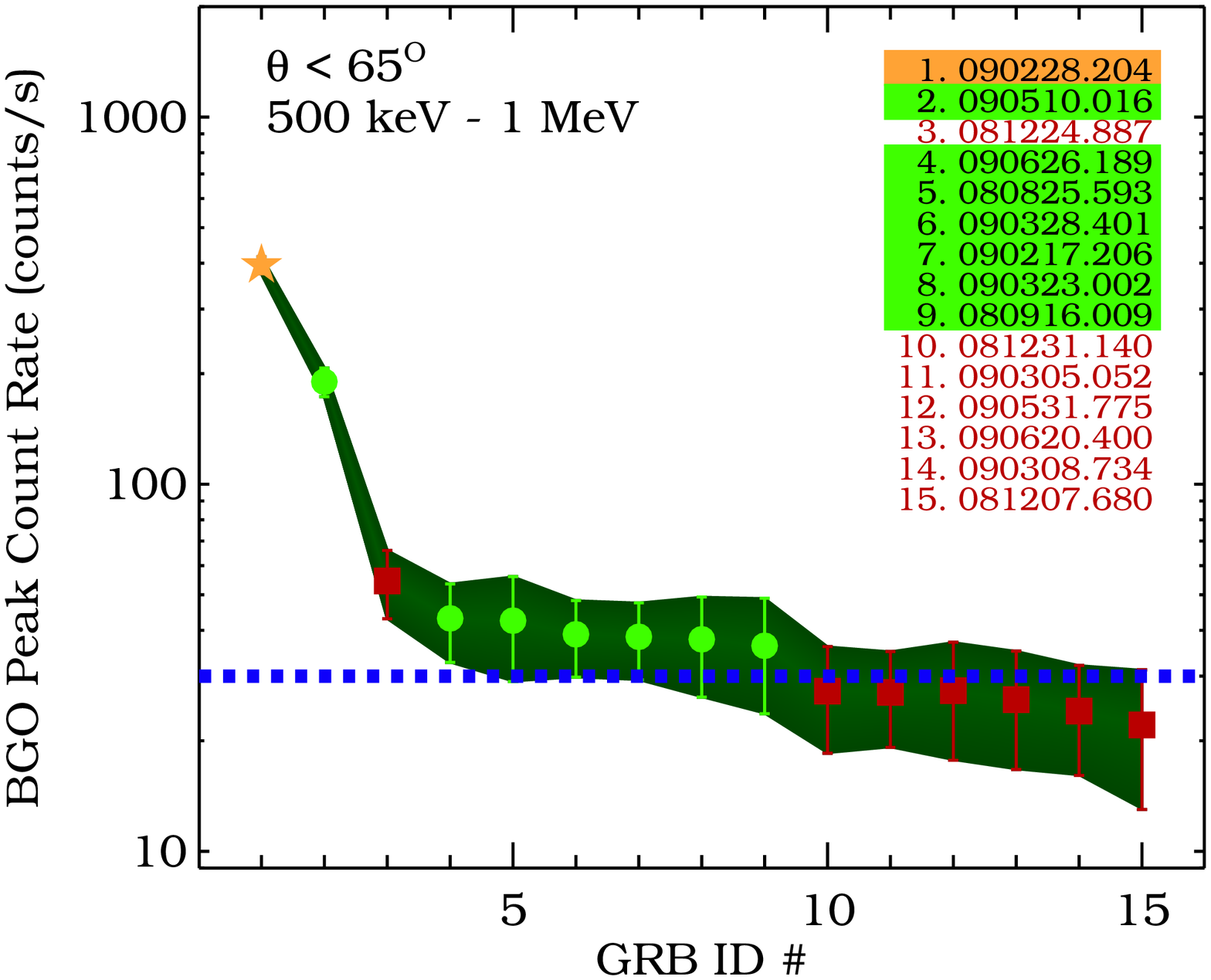} \\
\includegraphics[width=0.5\textwidth,bb=0 0 590 482,clip]{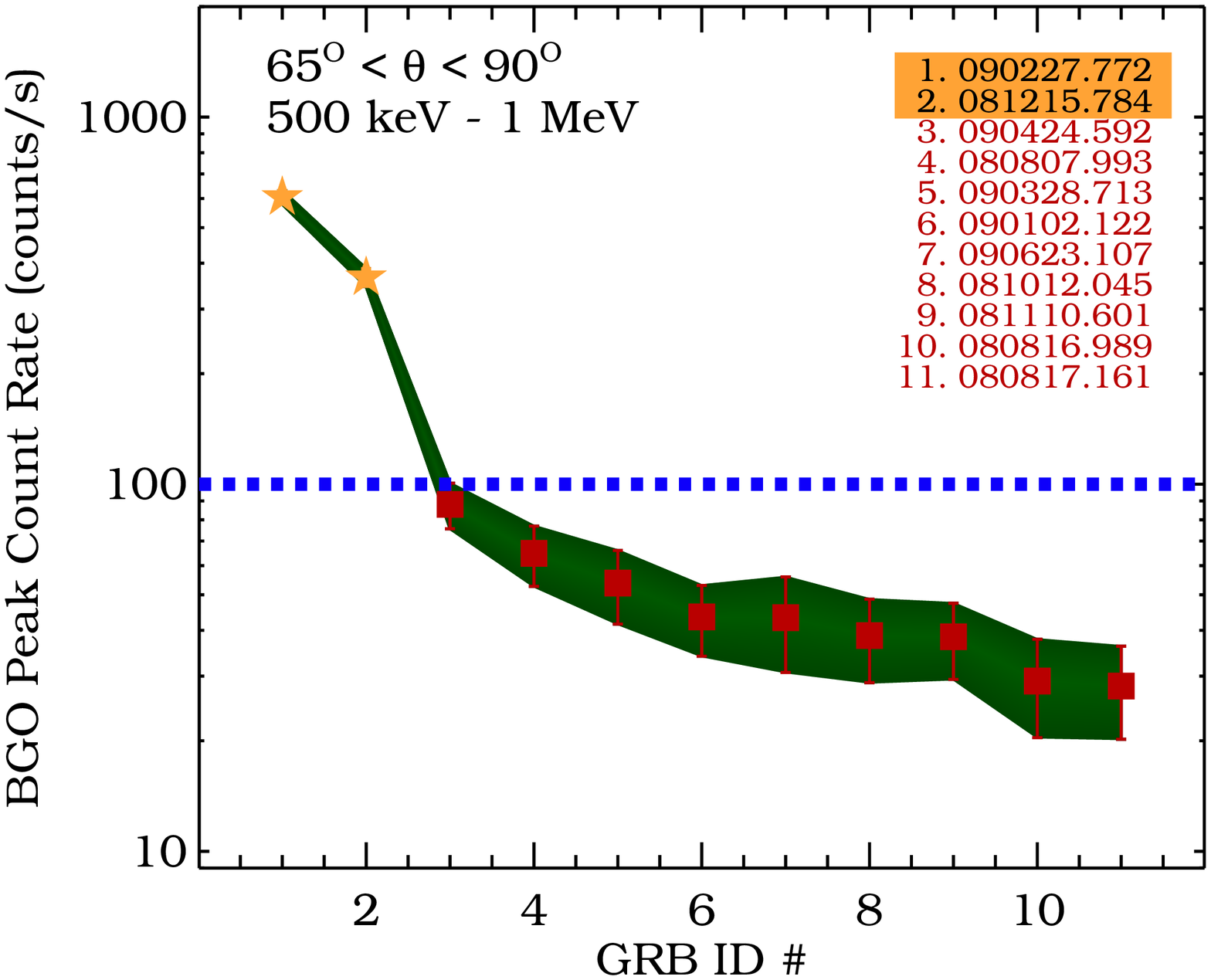}
\end{tabular}
\caption{BGO peak count rate measured in {\it Ch.1} (500 keV -- 1MeV)
for those bursts inside ({\it top panel})
and at the edge ({\it bottom panel}) of the LAT FoV.
Circles (green), stars (orange) and squares (red) represent
firm, marginal or no LAT detections, respectively. The dotted line
marks the arbitrary ``detection limit'' placed at 30 (top panel) and 100 (bottom panel) counts/s.
}
\label{Histo_PCR_LAT_boh}
\end{figure}

An interesting result of the BGO peak count--rate analysis
emerges by considering only those bursts,
which are located either inside or at the edge of the LAT FoV. 
(i.e. $\theta<$65\deg\ or 65\deg$<\theta< 90$\deg\ from 
the LAT boresight, respectively).
Figure \ref{Histo_PCR_LAT_boh} shows the BGO peak count rate
measured in {\it Ch.1} for 15 and 11 bursts, which
respectively fulfill these conditions. The GBM trigger IDs and numbers
for both subsamples are listed on the top right corner of each plot.
Circles (green), stars (orange) and squares (red) represent
firm, marginal or no LAT detections, respectively. The dotted line
marks a ``detection limit'', which was arbitrarily placed at 30 (top panel) and 100
cps (bottom panel) in the {\it Ch.1} peak count rate. For those bursts with
lower rates in {\it Ch.1} no detection has been yet reported from the LAT. 
The very promising \citet{PEL10}  LLE technique may recover the signal 
to confirm the LAT marginal detections and even  reveal undiscovered 
emission from some of the Figure \ref{Histo_PCR_LAT_boh} BGO bursts 
lying below these arbitrary thresholds.

The significance of the above analysis lies in its potential to provide 
a good predictor for LAT detections of GBM GRBs. We plan to implement 
the relevant software into the GBM FSW. Thus, starting from the burst 
location relative to the LAT FoV, the code would perform a finer 
computation of the BGO {\it Ch.1} count rate as measured between 
500\,keV and 1\,MeV. This information would then be sent to the ground, 
where it would be rapidly and automatically analyzed,  and subsequently 
(as the case maybe) provide a prompt  alert of space-- and ground--based 
GRB observatories.
\section{Temporal analysis}\label{Sec_Temp}
\subsection{Duration distributions}\label{durations}
\begin{figure}
\centering
\begin{tabular}{c}
\includegraphics[width=0.5\textwidth,bb=0 0 590 482,clip]{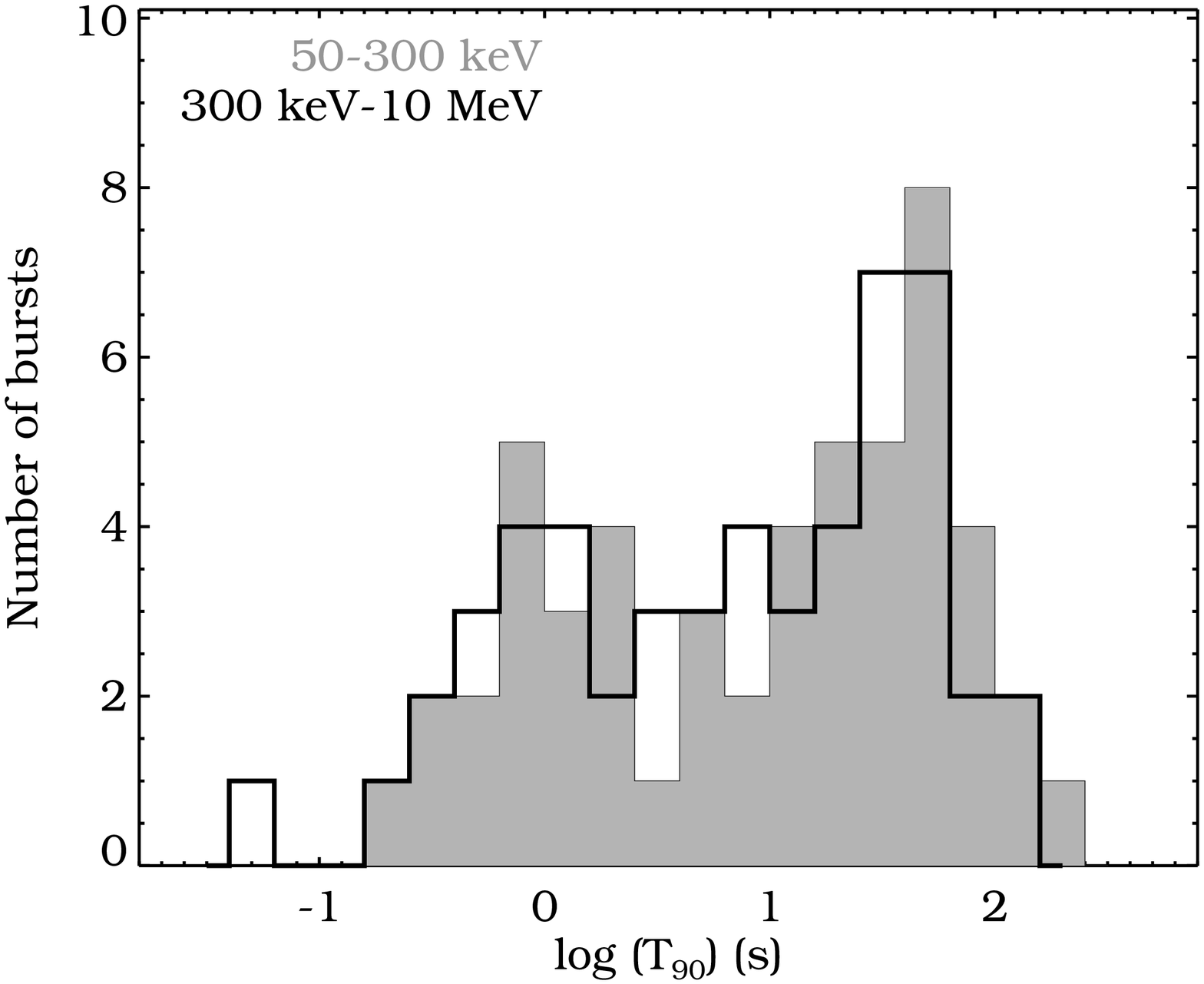} \\
\includegraphics[width=0.5\textwidth,bb=0 0 590 482,clip]{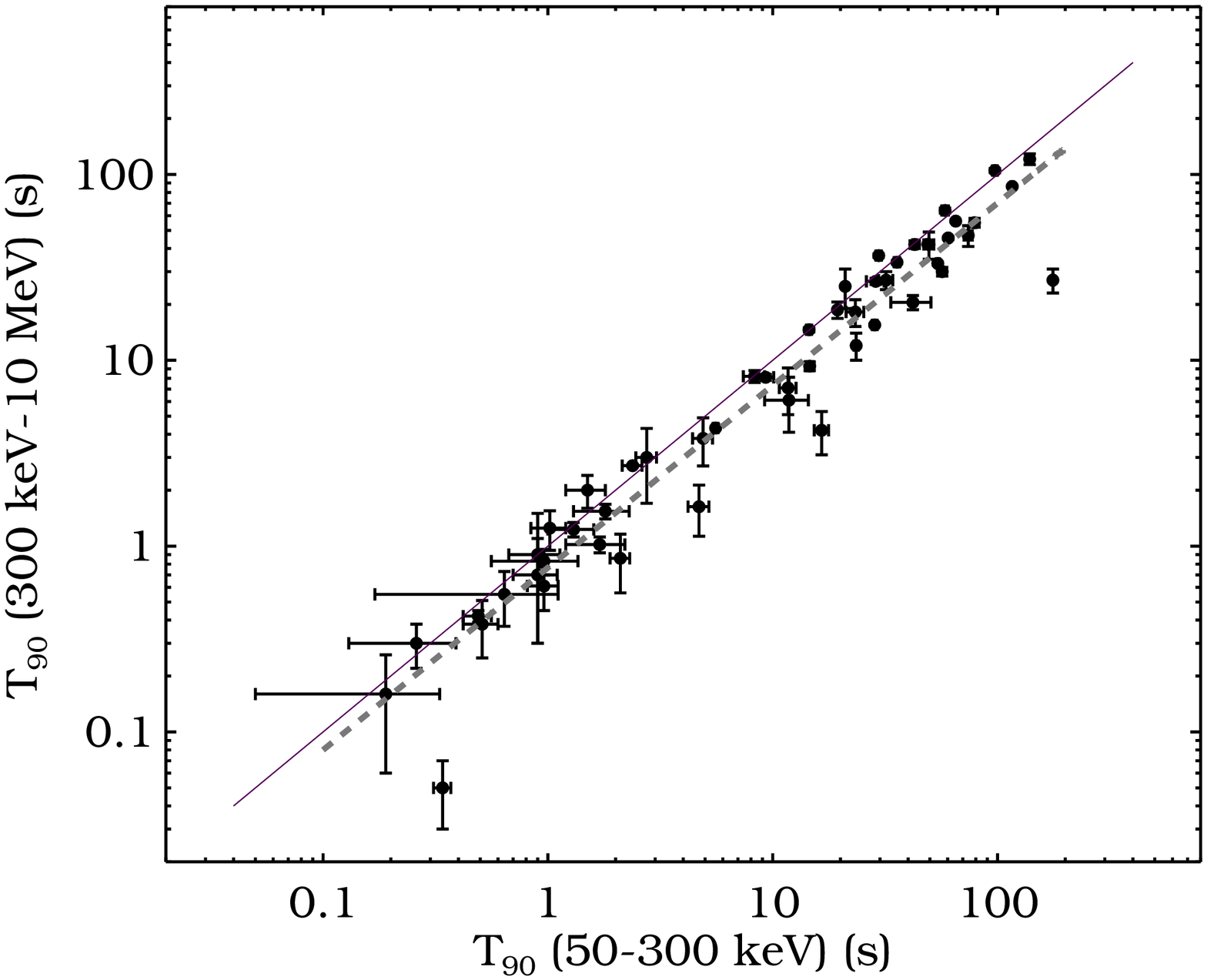}
\end{tabular}
\caption{{\it Top Panel}: Burst duration distributions in the 50--300\,keV ({\it gray filled histogram}) 
and in the 300 keV--10\,MeV ({\it black empty histogram}) energy ranges
for 52 bursts of our bright--burst sample. Both distributions
show a bimodal shape. {\it Bottom Panel}: Scatter plot
of the burst duration distributions. The continuous line represents
perfect linearity, while the dotted line represents the
fit applied to the data.}
\label{Plot_T90_Distributions}
\end{figure}

We used the CTIME data of the most illuminated
NaI detector for each GRB to compute the  \TN\ duration \citep{KOU93}
in the BATSE energy range of 50--300\,keV (for comparison reasons).
The background model was determined by fitting a low--order
($\leq$~4) polynomial function over at least a few hundred seconds 
before and after the trigger time. All standard \TN\ durations in the 50--300 keV energy range
were computed with the spectral analysis software RMfit\footnote{An ad--hoc version of RMFIT 
for GBM and LAT analysis was developed by the GBM Team and is publicly available at
{\tt http://fermi.gsfc.nasa.gov/ssc/data/analysis/}.} (version 3.8) 
\citep[see e.~g.,~][]{MAL05}, which was also employed for the spectral
analysis presented in \S \ref{Sec_Spec}. The resulting durations were thus estimated 
in photon space (i.e., the time it took to collect 90\%\ of the burst photons). 
A different approach (since the BGO data were not yet programmed in RMfit at that time) 
was adopted to compute the burst durations in the 300 keV--10 MeV energy range, 
where we estimated the \TN's in {\it count} space using software developed at MPE, 
thus obtaining a measure of the so--called ``BGO--duration''. Figure~\ref{Plot_T90_Distributions} 
includes both data sets for demonstration purposes. Although these are not directly comparable 
(photon {\it versus} count durations) they serve as qualitative trend indicators as we discuss below.

The initially selected sample of 52 GRBs includes 18 bursts with a duration 
\TN$<$ 3~s (50--300\,keV, {\it short GRBs}), and 34 bursts with a duration
\TN$>$ 3~s (50--300\,keV, {\it long GRBs}). 
The three longest bursts in the sample, with durations greater
than 100~s, are GRB~081009, GRB~090323 and GRB~096018, while the 
three shortest ones are GRB~081226B, GRB~090328B and GRB~090228.
The top panel of Figure~\ref{Plot_T90_Distributions} shows the distribution of \TN, calculated 
in the 50--300\,keV and in the 300 keV--10\,MeV energy ranges, for all 52 bursts.
We note that the duration bimodality found in the BATSE data \citep{KOU93} 
is evident even in this small sample of GBM data and 
in both count and photon -- space durations. We fitted 
each distribution with a double Gaussian function to estimate the medians and deviations. We found that 
short bursts peak at 1.2 $\pm$0.3\,s (1.04 $\pm$0.16\,s) in the 50--300\,keV (300\,keV--10\,MeV)
bands, while long bursts peak at 33 $\pm$5 \,s (25 $\pm$8\,s), respectively.
In the bottom panel of Figure \ref{Plot_T90_Distributions} we plot the
\TN\ distributions against each other. The dashed line represents
a linear fit applied to the data, It has a slope of $0.98 \pm0.03$,
and lies below the bisector (continuous line), indicating that on the average, the BGO durations are smaller than 
the NaI's. This result agrees qualitatively with the earlier 
findings of \citet{RIC96} that GRB durations decline with energy. 
In \S \ref{duration_evolution} we explore 
and expand this relationship to the MeV range for the first time.
\subsection{Evolution of Duration with Energy}\label{duration_evolution}
We followed the approach described by \citet{RIC96}, who presented the
analysis of 72 intense GRBs from the BATSE 3B catalog.
They measured their \TN\ in four broad 
energy channels, namely 25--50\,keV, 50--100\,keV, 100--300\,keV, and 
$>$300\,keV. The bursts used for
their study were uniformly selected by their peak photon 
flux on the 64~ms time scale. Thanks to the broader
BGO energy coverage, we can measure \TN\ over the five 
additional energy channels described in \S\ \ref{Sec_Samp_Selec} (the only exception 
is {\it Ch.0}, which is here defined between 300--500\,keV, to match 
the upper edge of the BATSE \TN\ values, for comparison reasons).
The \TN\ calculated using NaI only data in the 50--300\,keV
energy band represents our lowest measurement; this energy interval was not further subdivided.

The energy intervals used for our analysis are listed in the first column
of Table \ref{Tab_Dur_En}. Columns 2--4 give the number of bursts
detected over the different energy intervals, where T
indicates the total number of GRBs, and L and S indicate
long and short bursts, respectively.
The average \TN\ values calculated using NaI only data in the 50--300\,keV
for each subgroup (again in photon space) are given in columns 5 and 6.
It is worth noting that {\it the bursts detected up to
higher energies are systematically longer}. This behavior is seen
for both long and short GRBs.

\begin{figure}[b!]
\centering
\begin{tabular}{c}
\includegraphics[width=0.5\textwidth,bb=0 0 590 482,clip]{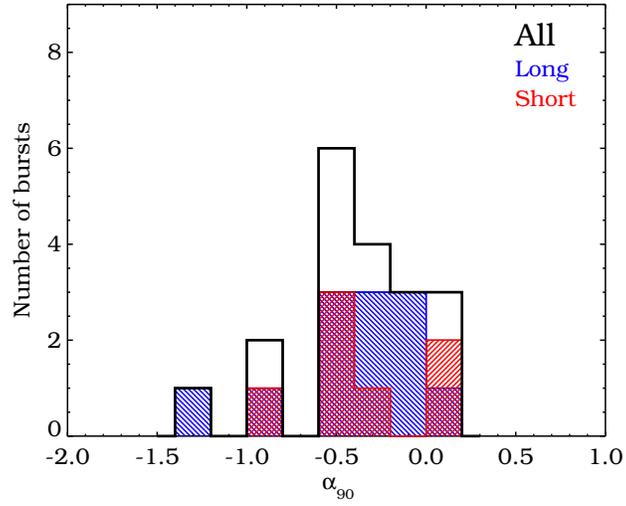}
\end{tabular}
\caption{Distribution of the best fit indexes $\alpha_{90}$ of power--law 
fits to the burst durations over different energy channels.
{\it Blue} and {\it red histograms} represent the distributions
of long and short GRBs, respectively. The black histogram 
represents the entire sample.}
\label{Histogram_alpha}
\end{figure}

We then fitted the energy {\it versus} duration values for 
each burst in the sample, excluding the NaI data points 
(to ensure a homogeneous set) to a power--law function given by
\begin{equation}
T_{90} = A_{90} \, E^{\alpha_{90}}
\end{equation}
using a $\chi^2$ minimization technique. We used the central
energy value to represent each energy channel in the fit.
The distribution of the best fit indexes, $\alpha_{90}$, 
over all energy intervals is shown in Figure \ref{Histogram_alpha}.
Blue and red shaded histograms represent the distributions
of long and short GRBs, respectively.
The average value of $\alpha_{90}$ calculated with this
technique is $-0.39\,\pm\,$0.04\,(0.34), 
where the first error is the error in the mean, 
and the error given in parenthesis is the standard deviation 
of the distribution. This value is in good agreement with the result 
of \citet{RIC96} and indicates that the same overall trend of declining 
duration {\it versus} energy in the GRB light curves continues 
up to at least 10\,MeV. 

\begin{figure}
\centering
\begin{tabular}{cc}
\includegraphics[width=0.48\textwidth,bb= 0 0 518 482,clip]{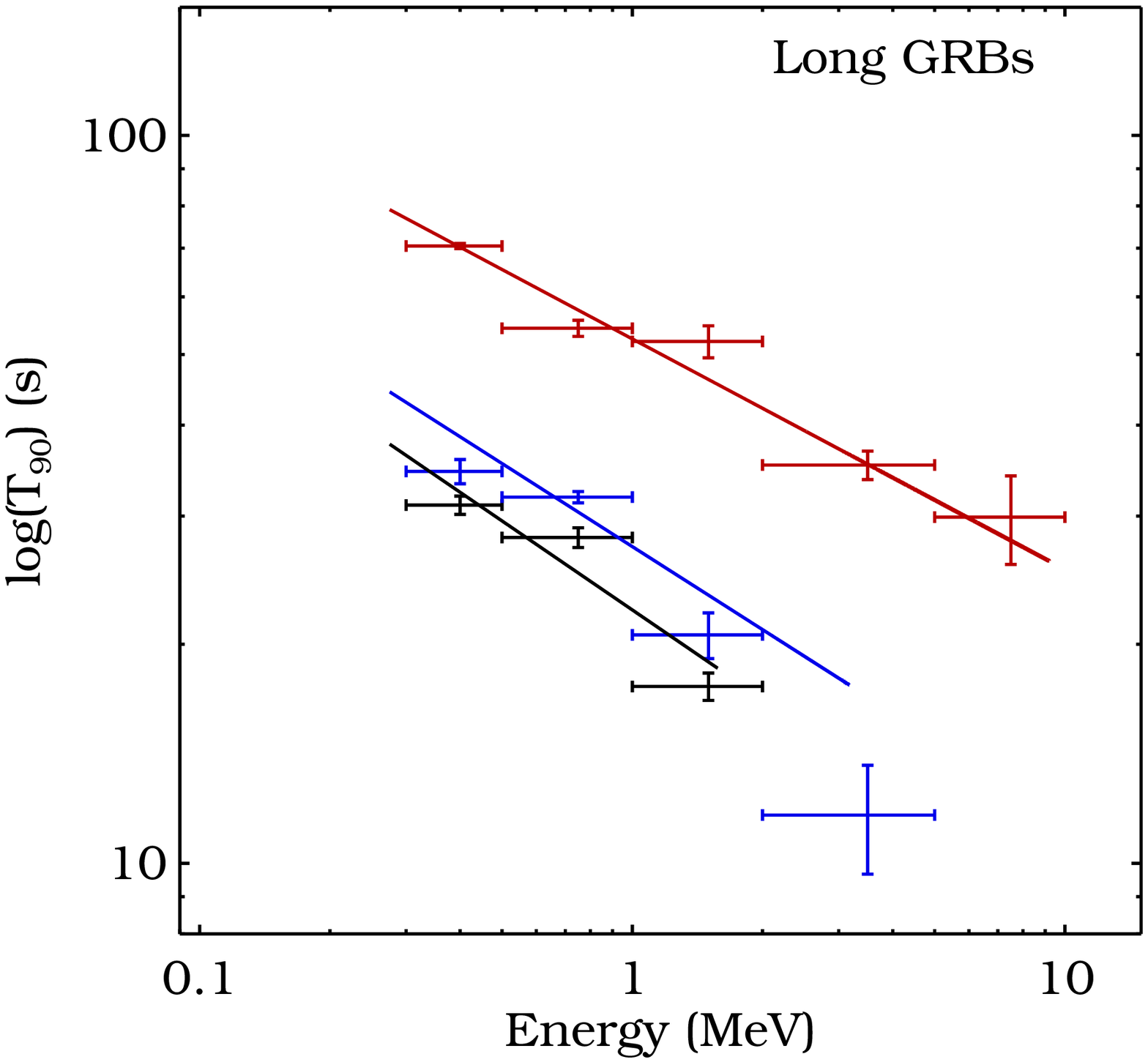}
\includegraphics[width=0.48\textwidth,bb= 0 0 518 482,clip]{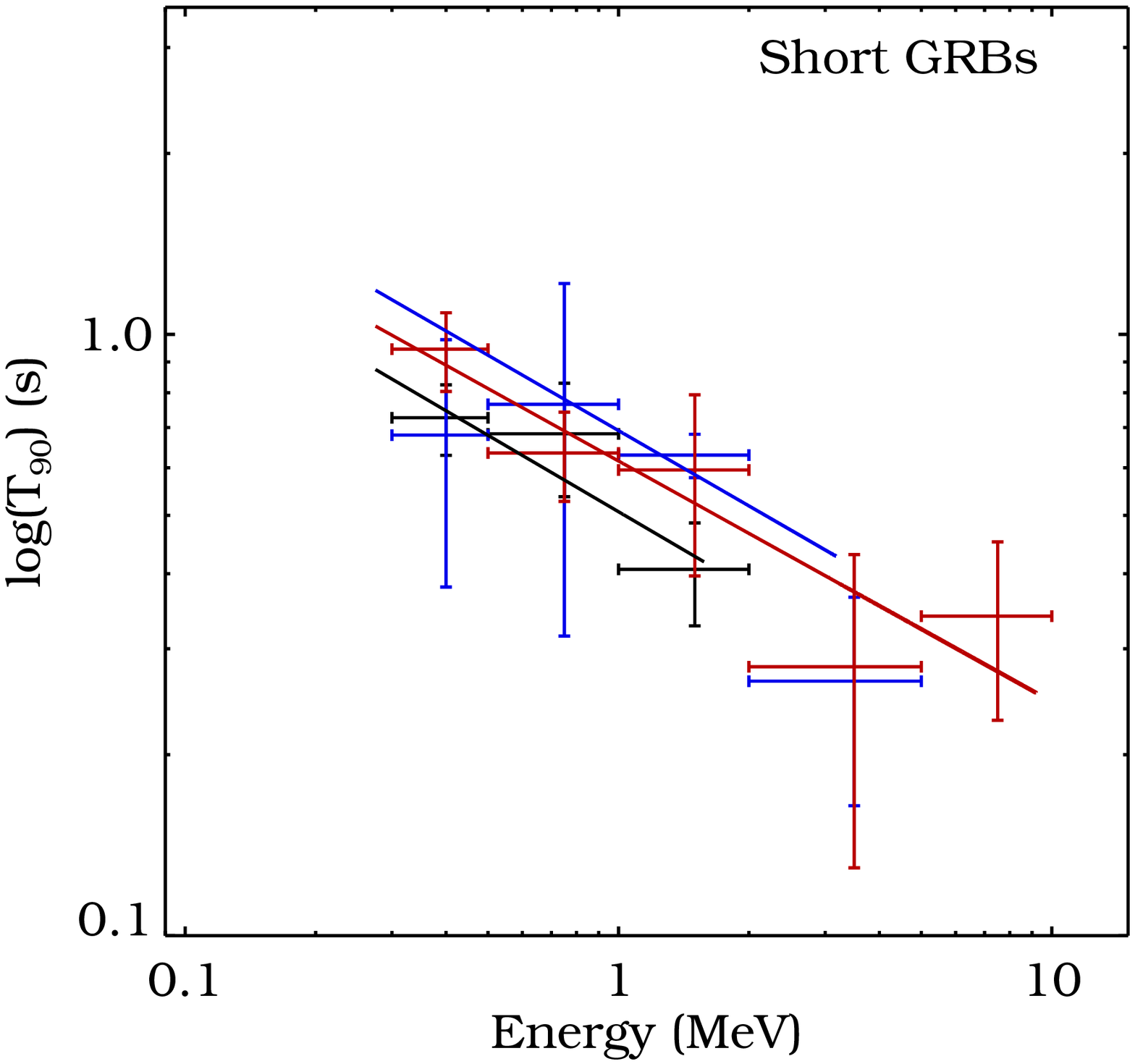}
\end{tabular}
\caption{Evolution of the \TN\ duration with respect to energy for
four subsamples of 12 long GRBs (left panel) and 7 short GRBs (right panel). 
Energy bands are over the 5 BGO CTIME channels. Different curves correspond
to different burst sets for which such duration measurements were possible
(see also text and Table \ref{Tab_Dur_En}).}
\label{Plot_Duration_Energy}
\end{figure}

We now proceed to explore the duration--energy relation in subsamples 
of our 52 GRBs selected according to their highest detection in each 
of the 5 BGO channels described above.  The sample sizes are defined in 
Table \ref{Tab_Dur_En}; e.g., 22 long GRBs are detected up to {\it Ch.1} 
and a mere 4 up to {\it Ch.5}.  For each subsample, we first 
average the values of \TN\ for each energy channel, and then 
fit a power law to these data as we did in the total sample. 
These fits were computed separately for long and short bursts and 
are shown in two panels in Figure~\ref{Plot_Duration_Energy}.
The values of their $\alpha_{90}$ are given in columns 7 and 8 of Table \ref{Tab_Dur_En}. 
The left panel of Figure \ref{Plot_Duration_Energy} clearly exhibits the trend for the 
longest subsample to extend to higher energies; we can see the same 
trend in the short GRB subsample albeit with lesser statistical 
significance. Whether this trend is simply a selection effect 
or an intrinsic GRB property remains to be determined with a 
larger GRB sample, including events with LAT detections possibly 
extending this relation to higher energies. We find evidence 
for slight curvature of \TN\  {\it versus} Energy in both duration 
modes indicating a potential cut off. The detailed study of these 
relations using a uniform data set and time--resolved binning 
of the NaI durations is the subject of another publication, 
currently in preparation \citep{BIS11}. We note here 
that the red subsamples include GRBs that have indeed been 
detected with the LAT in GeV energies. 
\section{Spectral Analysis}\label{Sec_Spec}
During the BATSE era, GRB spectra were well represented by 
a broken power--law \citep[the empirical {\it Band} 
function,][]{BAN93} in the BATSE energy band (K06). 
In our spectral analysis, we fitted
the {\it time--integrated} spectra of the 52 bright 
bursts in Table~\ref{Tab_Prop_All} with two functions: 
(i) the {\it Band} function, and (ii) the {\it Comptonized (Comp)} 
model. The second represents a low--energy power law with an 
exponential high--energy cutoff, which is equivalent to the 
{\it Band} function without a high--energy power law 
(i.~e., $\beta \rightarrow$ $-\infty$). More details 
regarding both spectral functions can be found in K06.
Since we excluded faint or soft GRBs from our sample, the simple 
power--law model was not used. Moreover, no additional extra--components
were fitted to both spectral models. 
For a detailed time--integrated spectral analysis of a subsample 
of three short GRBs exploring several multi--component emission models see \citet{GUI10}. 
\citet{GUI11} have also identified an additional spectral component in the 
time--integrated analysis of GRB~100724B; the appearance of these 
components in the high--energy part of the GBM spectrum could be 
another predictor for the bursts expected to be seen with the LAT.

We performed broadband spectroscopy simultaneously on the $\sim$ 8~keV to 
1~MeV NaI data and the $\sim$ 250~keV to $\sim$38~MeV BGO data.  
In some cases an effective area correction had to be fit to the 
BGO data \citep[see e.~g.][]{ABD09b} to match the model normalizations given 
by the NaI data; this correction is usually consistent 
with the uncertainties in the GBM detector responses ($<$10\%).
Columns 8 and 9 of Table~\ref{Tab_Prop_All} list the time 
intervals used for each burst. All times are referred to 
the burst trigger time $T_0$.

The time--integrated spectral best fit results are presented in Table~\ref{Tab_Spec_MAIN} for
each burst. The GRB trigger number is given 
in column 1, while the best model for the spectral fit is listed in
column 2. The spectral parameters are given in columns 3--7.
Column 8 lists the effective area correction factors.
The quality of fit in terms of CSTAT\footnote{The
CSTAT statistics is equivalent to the XSPEC \citep{ARN96}
implementation of the Cash statistic \citep{CAS79}.} over 
degrees--of--freedom (DOF) is listed in column 9.
The {\it Band} function was preferred for those bursts
for which an improvement of $>$~10 was observed in
the CSTAT statistic over the {\it Comp} function.
This assures that the spectra have a well--identifiable 
high--energy power--law component. 
We find that the {\it Band} function is preferred over the {\it Comp} model 
in 27 out of 52 cases (52 \%).
Most short bursts are best fit by a {\it Comp}
model: Only 5 out of 18 short GRBs are 
best fit by the {\it Band} function. Three of those are 
the brightest bursts in our sample, for which \citet{GUI10} have 
reported the presence of an extra power--law component in the GBM data. 
We proceed below in the description 
of the spectral parameter distributions, their correlations 
and their comparisons with empirical relationships in the literature.  
\subsection{Distributions of Spectral Parameters}
\begin{figure}
\centering
\begin{tabular}{c}
\includegraphics[width=0.48\textwidth,bb=0 0 590 482,clip]{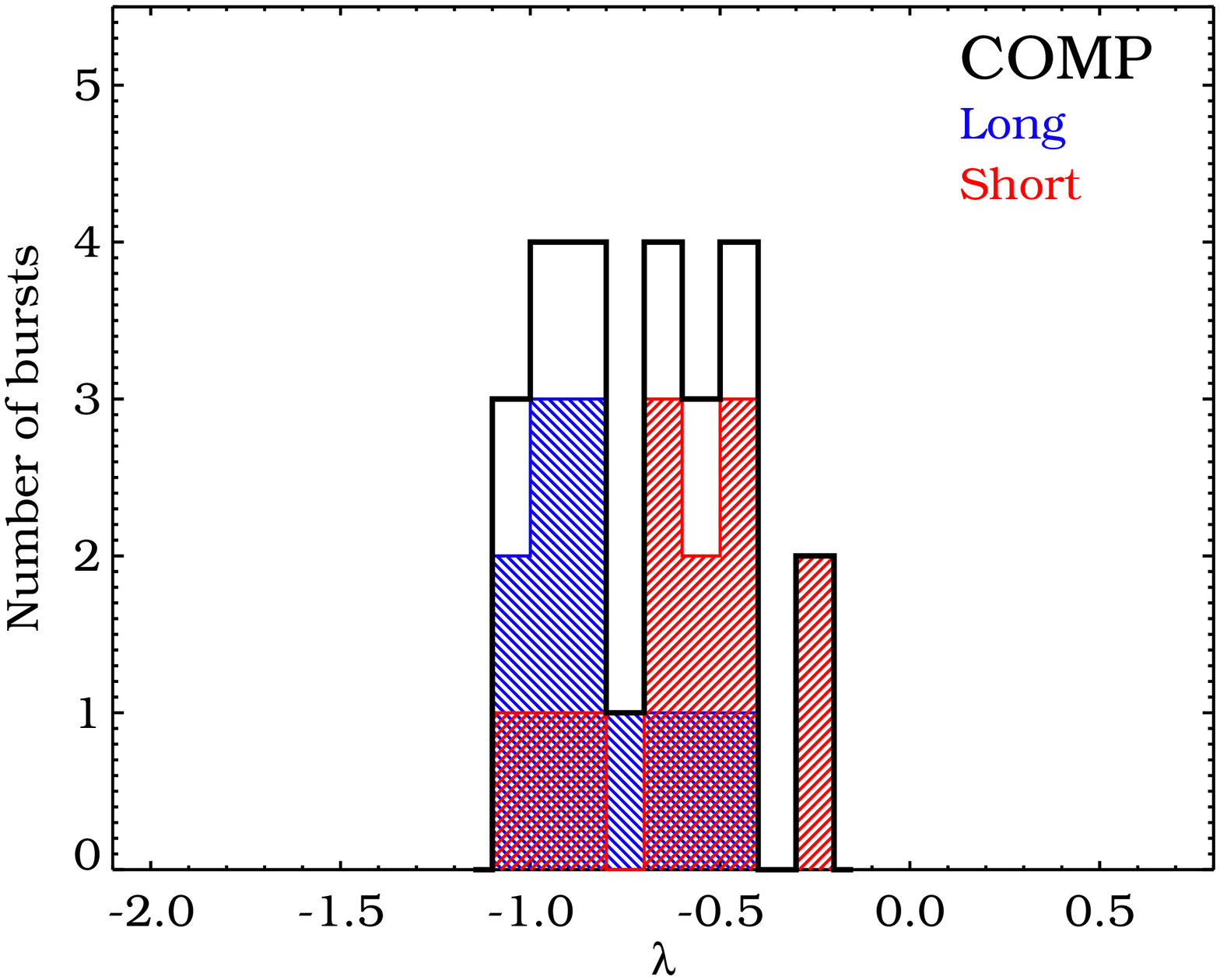} \\
\includegraphics[width=0.48\textwidth,bb=0 0 590 482,clip]{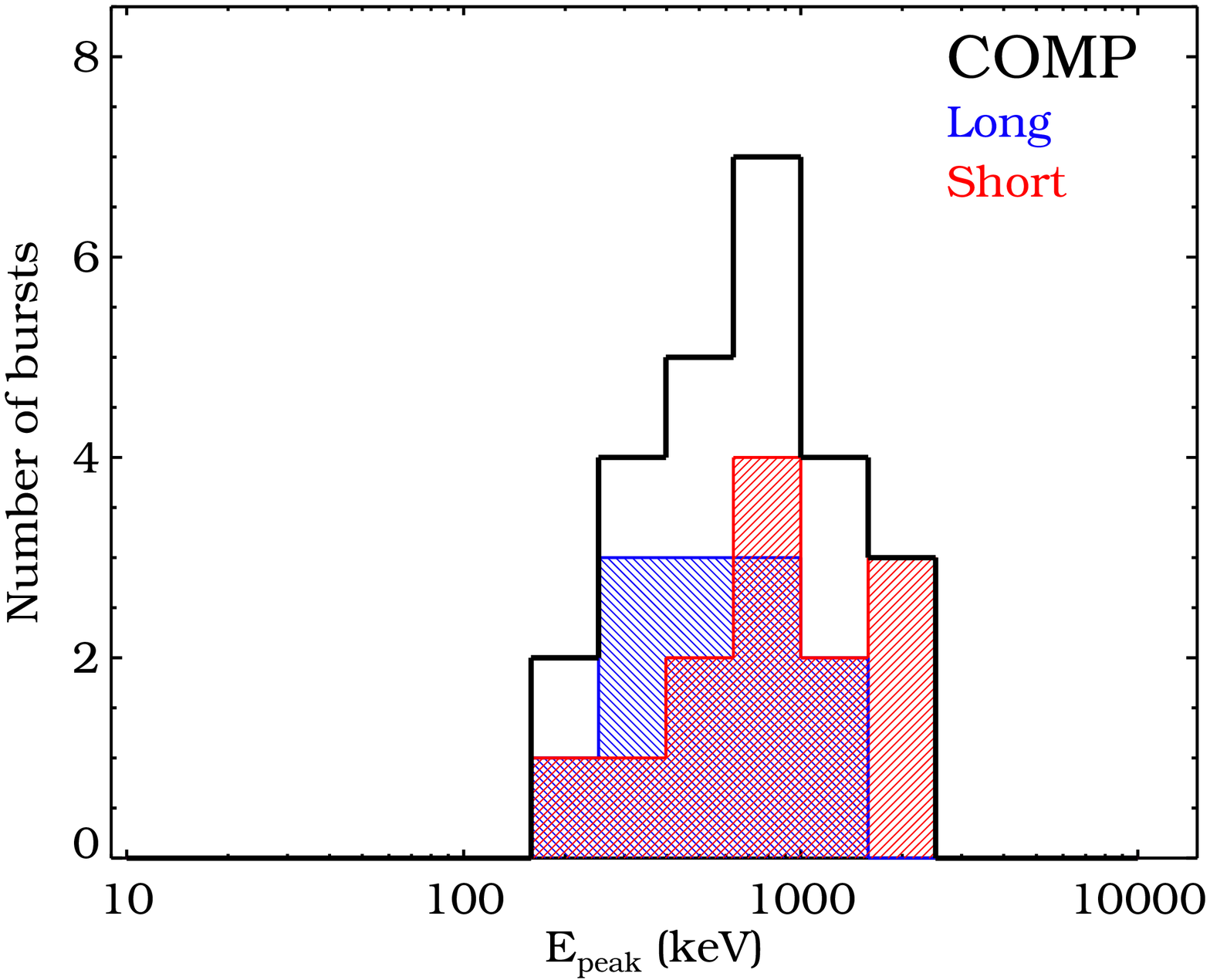}
\end{tabular}
\caption{Index $\lambda$ ({\it top panel}) and \Epeak\ ({\it bottom panel}) distribution of the time--integrated
spectra which are best fitted with the {\it Comp} model.
{\it Blue} and {\it red histograms} represent the distributions
of long and short GRBs, respectively.}
\label{Histo_All_Param_COMP}
\end{figure}

\begin{figure}
\centering
\begin{tabular}{c}
\includegraphics[width=0.45\textwidth,bb=0 0 590 482,clip]{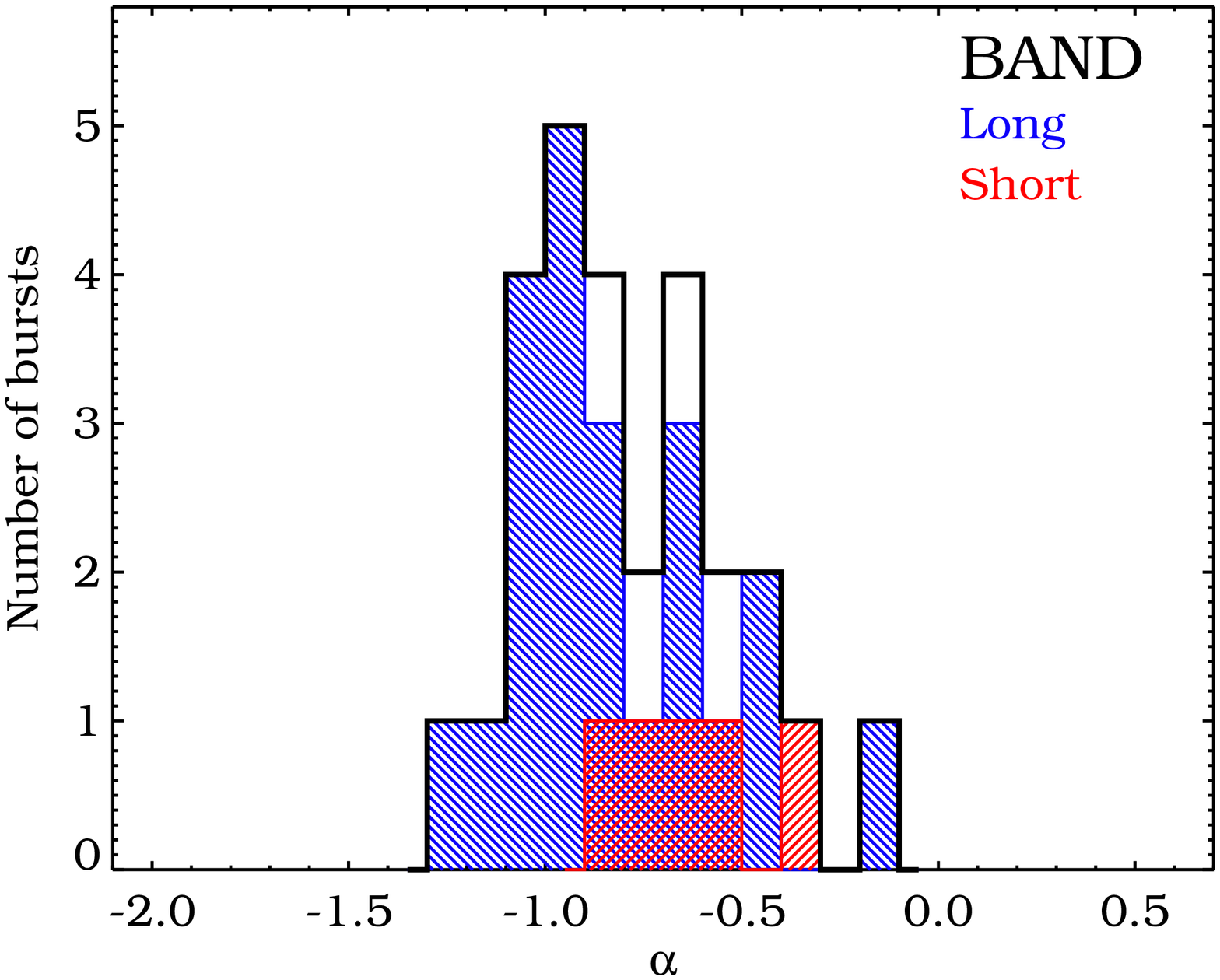} \\
\includegraphics[width=0.45\textwidth,bb=0 0 590 482,clip]{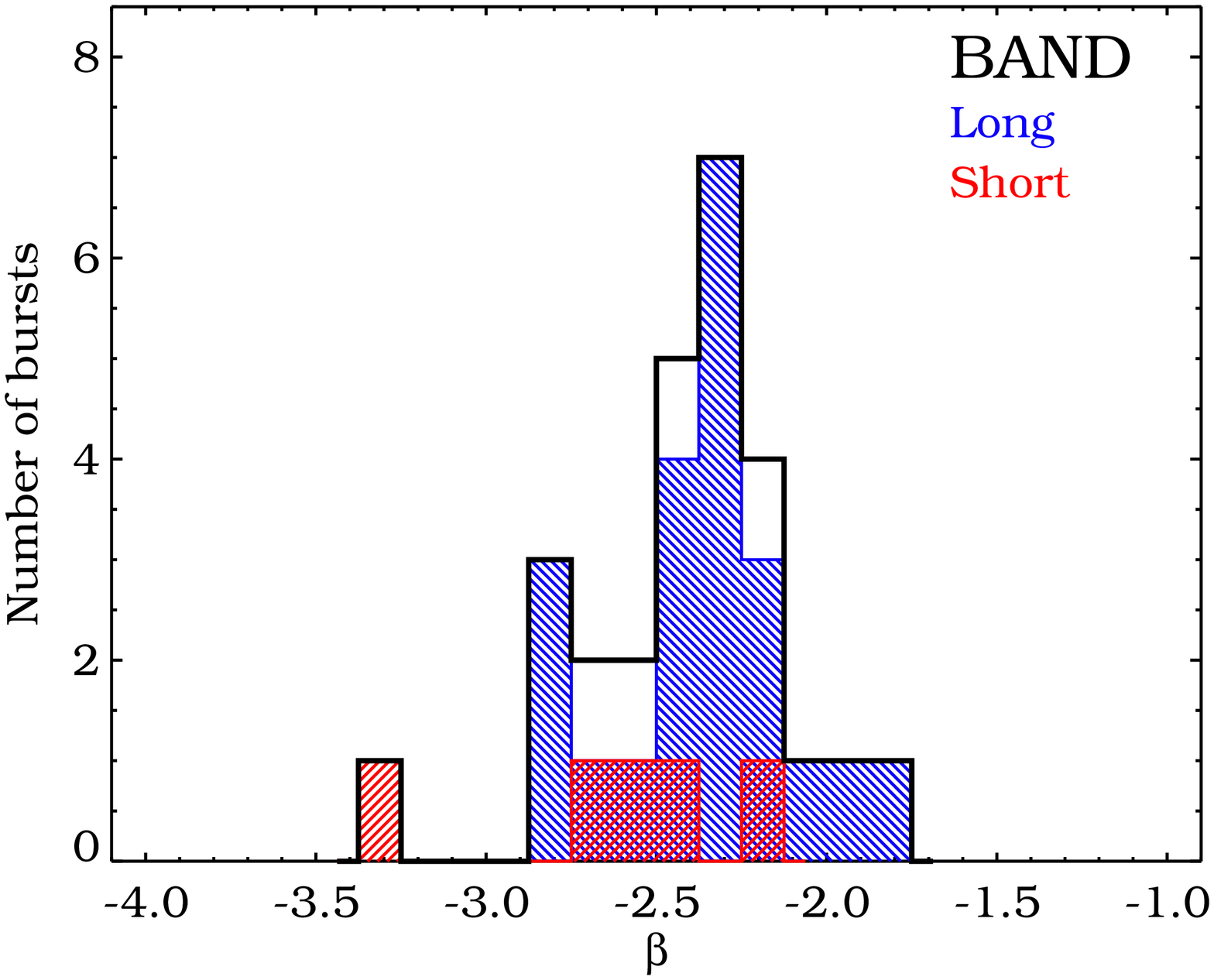} \\
\includegraphics[width=0.45\textwidth,bb=0 0 590 482,clip]{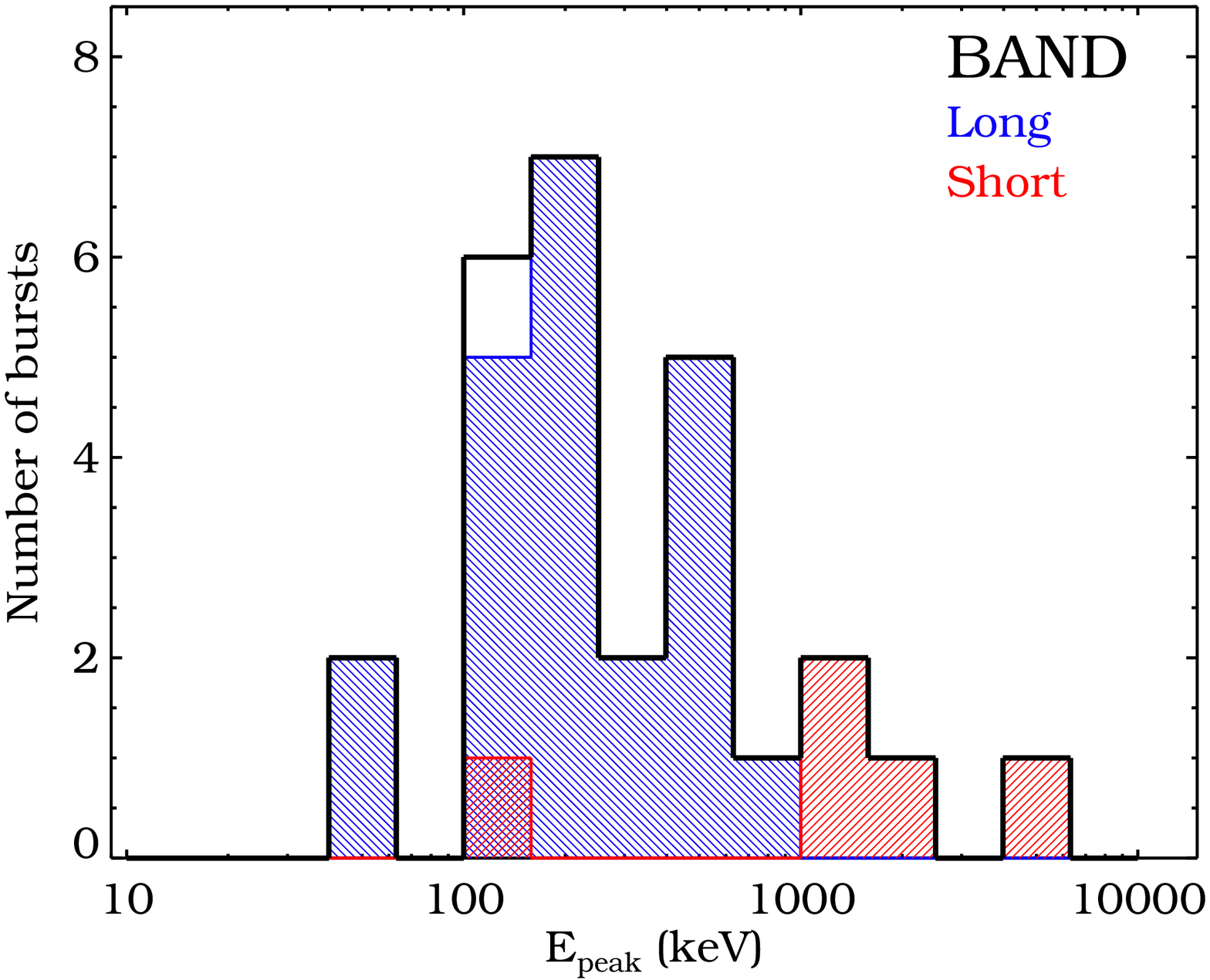}
\end{tabular}
\caption{Low--energy index $\alpha$, high--energy index $\beta$
(top panels) and \Epeak\ (bottom panel) distribution of the time--integrated
spectra best fitted with the {\it Band} function.
Blue and red histograms represent the distributions
of long and short GRBs, respectively.}
\label{Histo_All_Param_BAND}
\end{figure}

The distribution of the two parameters of the {\it Comp} 
model (index $\lambda$ and \Epeak) are shown in Figure
\ref{Histo_All_Param_COMP}; those of the {\it Band} function 
parameters are displayed in Figure~\ref{Histo_All_Param_BAND}.  
In both figures the \Epeak\ parameter distribution is plotted at the bottom. 
In Figure~\ref{Histo_All_Param_BAND}, the two top panels show
the low--energy index $\alpha$ and
the high--energy index $\beta$. In all plots of the following 
sections, blue, red, and black histograms represent the 
distributions of 34 long, 18 short, and the entire sample 
of 52 bright GRBs, respectively. 

The spectral parameter distributions show that short bursts 
tend to have larger  $\alpha$ ($\lambda$) and smaller $\beta$ 
values. The four short bursts best fitted with a {\it Band} 
function show higher \Epeak\ than long bursts (Figure~\ref{Histo_All_Param_BAND}, 
bottom panel); in three of them we found a value of $\beta<-$2.6. 
The {\it Comp} model is preferred for the other 13 short bursts. 
The {\it Comp} \Epeak\ distribution of the total sample peaks at 
$\sim800$\,keV, while the {\it Band} \Epeak\ is much lower, 
around 200\,keV. Below we compare our results with those of K06. 
\subsection{Comparison to the BATSE bright \\ burst catalog}\label{Sec_Comp_BATSE}
\begin{figure}
\centering
\begin{tabular}{cc}
\includegraphics[width=0.45\textwidth,bb=0 0 590 482,clip]{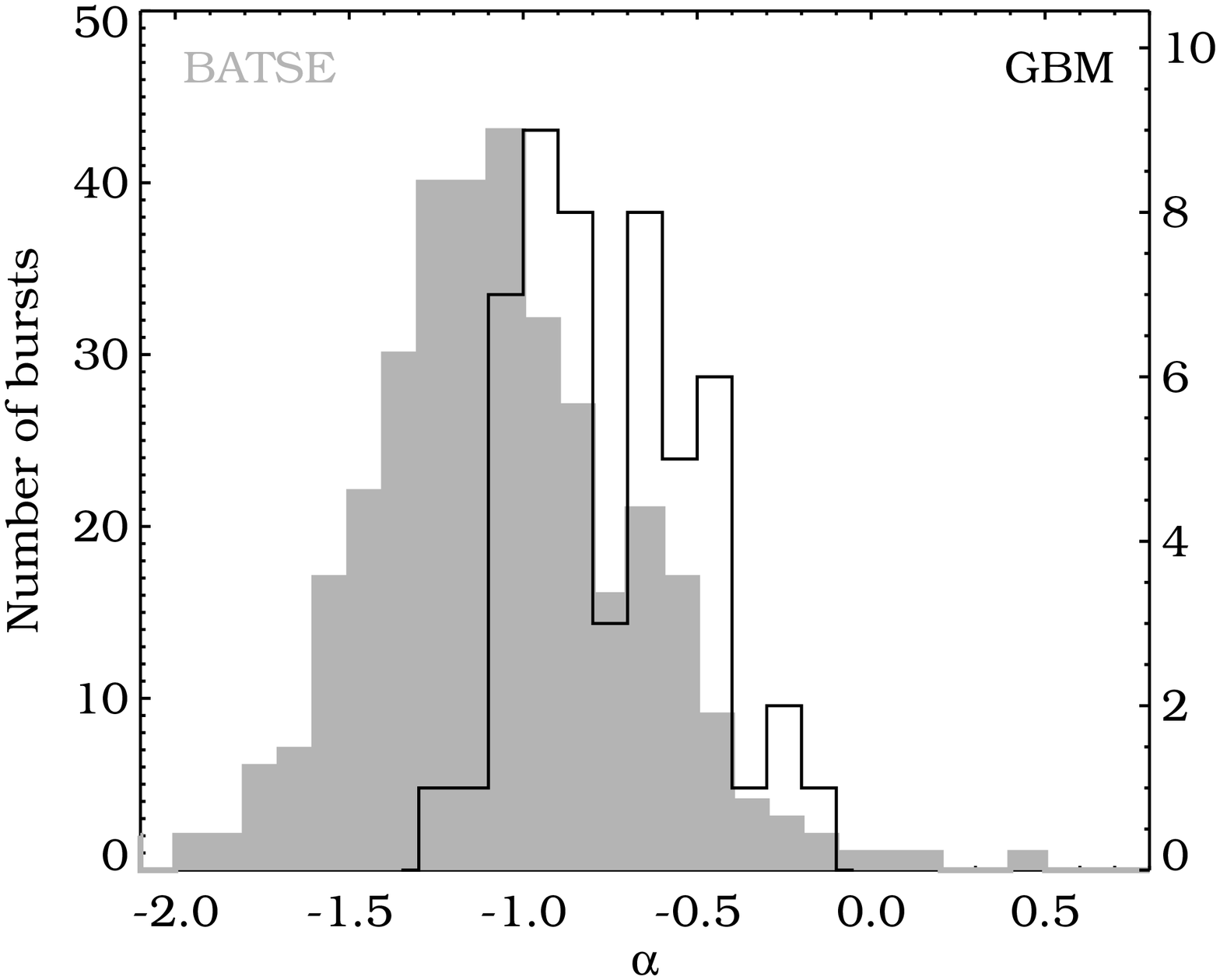}  & 
\includegraphics[width=0.45\textwidth,bb=0 0 590 482,clip]{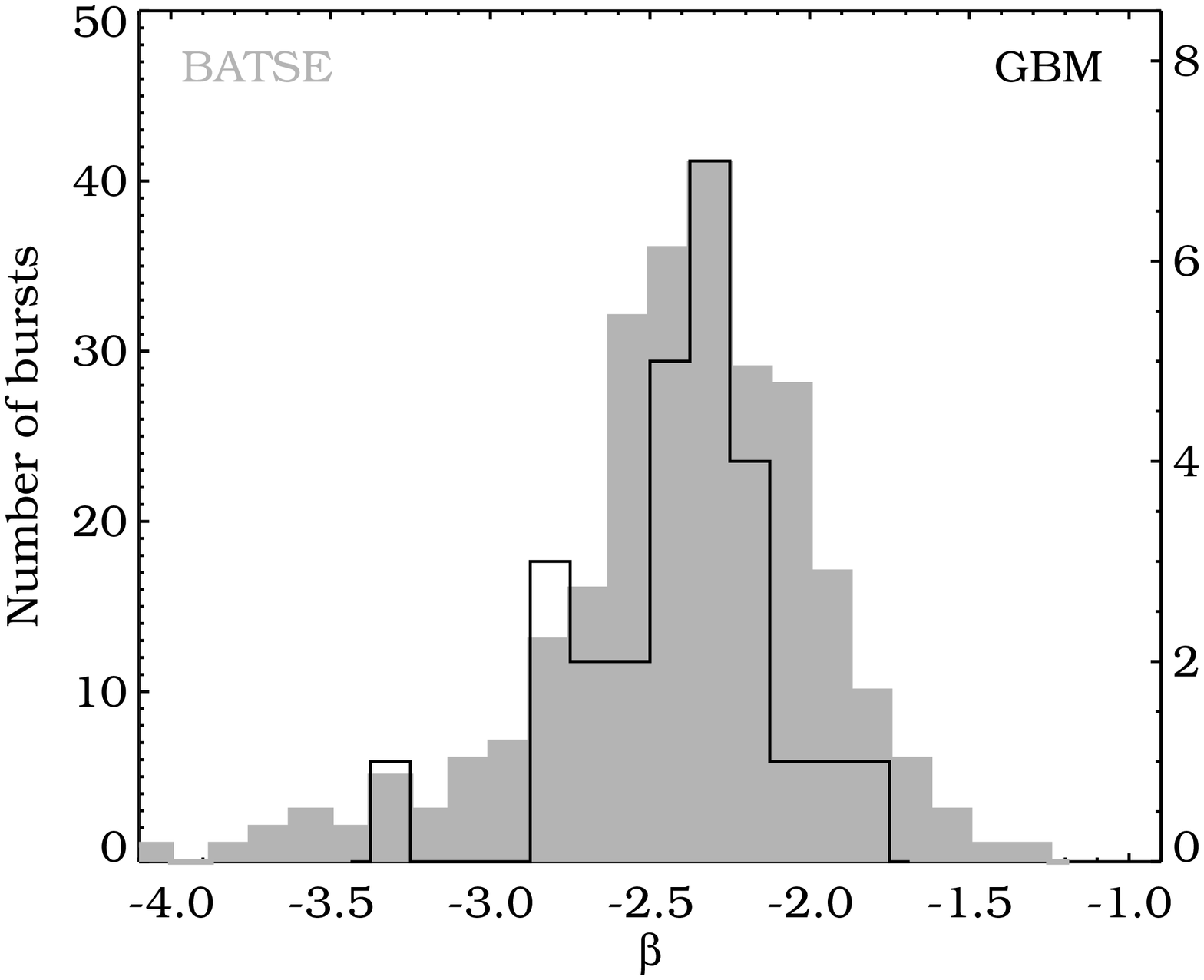} \\
\includegraphics[width=0.45\textwidth,bb=0 0 590 482,clip]{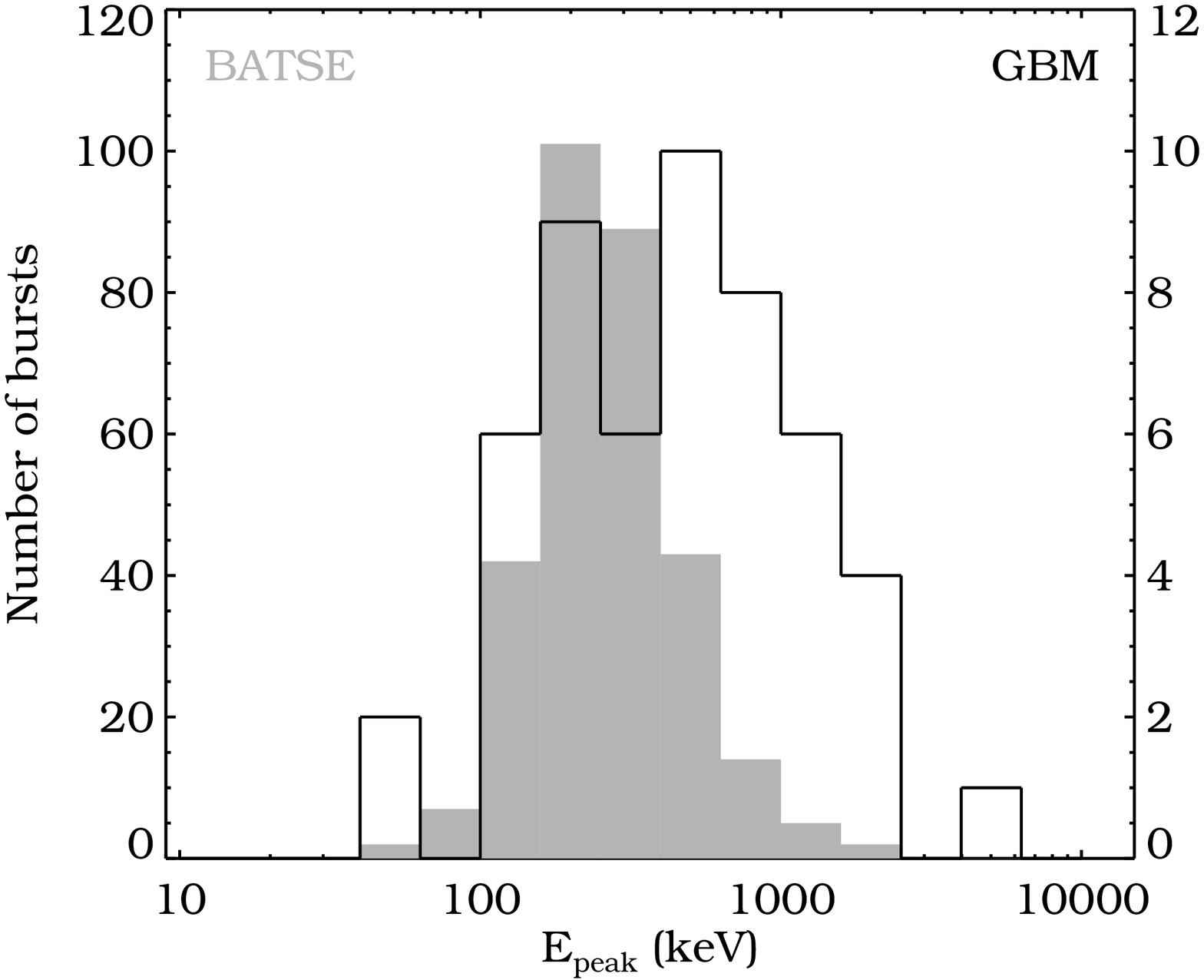}   &
\includegraphics[width=0.45\textwidth,bb=0 0 590 482,clip]{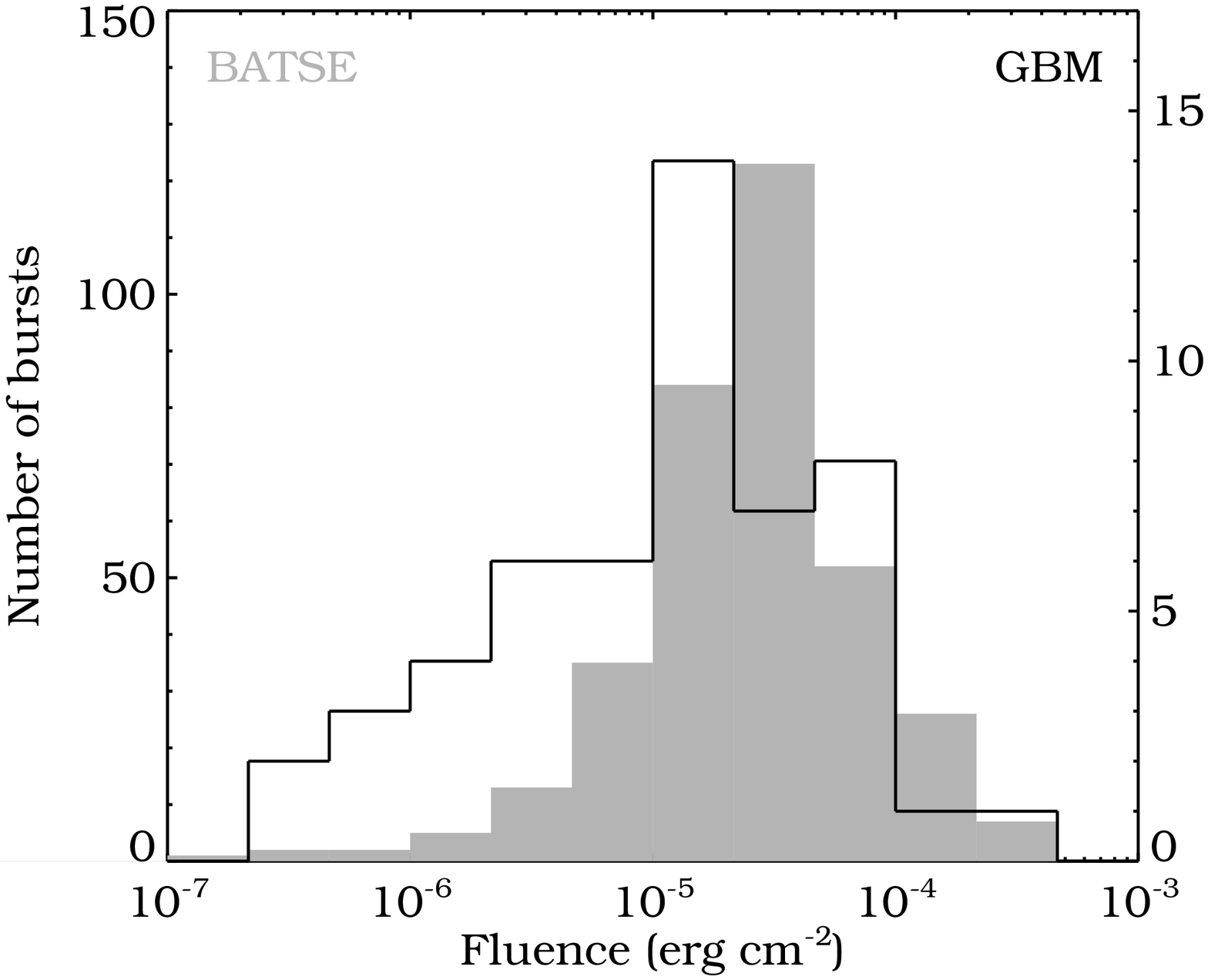}
\end{tabular}
\caption{Low--energy index $\alpha$ (top left panel), high--energy index $\beta$
(bottom left panel), \Epeak\ (top right panel) and energy fluence (bottom 
right panel) distributions of time--integrated
spectra from 350 bright BATSE bursts (grey filled histogram) from K06 (BEST sample) 
and 52 bright GBM bursts (black histogram). The GBM energy fluence is computed in the
standard BATSE energy range, namely between 25 and 2000\,keV. 
The GBM parameter distributions follow the right $y$--axis.}
\label{Histo_All_Param_BATSE}
\end{figure}

We compare the spectral parameter distributions of
our 52 bright GBM bursts with the 
distributions of bursts from the BEST\footnote{\citet{KAN06}
designated the model with the more constrained 
parameters as the best--fit (BEST) model.}
sample of K06.
Our sample is much smaller than the one recently used by 
\citet{NAV11}, which comprises the entire GBM GRB database. 
The K06 spectral catalog comprises 350 bright GRBs observed 
with BATSE ($\sim$20~keV--2~MeV) and is the most comprehensive 
study of spectral properties of GRB prompt emission to date, 
thus representing a perfect sample for comparing with GBM burst properties.
Comparison histograms of GBM and BATSE low--energy ($\alpha$), high--energy ($\beta$)
and \Epeak\ spectral parameter distributions are shown in 
Figure~\ref{Histo_All_Param_BATSE}.
The GBM distributions (black empty histograms) are overplotted 
on the BATSE ones (gray filled histograms). The former follow the
right $y$--axis and are rescaled for comparison purposes. 

While the two $\beta$ distributions look similar, differences appear
in the $\alpha$ and \Epeak\ distributions. After
one year of operations, GBM detected a sample
of bright bursts which tend to have 
larger $\alpha$ and higher \Epeak\ values 
than what was observed in 10 years with the BATSE instrument.

The energy fluence distribution for both samples is calculated 
in the (BATSE) energy range of 25--2000\,keV and is also shown 
in the bottom right panel of Figure~\ref{Histo_All_Param_BATSE}. 
It becomes immediately
evident that the two samples were selected following different criteria. 
The K06 sample comprises more GRBs with higher fluence than the GBM sample. 
This difference is even more clearly demonstrated in Figure~\ref{Plot_Epeak_Fluence}, 
where we plot the GBM burst \TN\ durations versus their fluences 
(top right panel) and peak fluxes (calculated over 128--ms, bottom right panel) 
between 8\,keV--40\,MeV. 
A comparison of the two panels shows that a high--fluence criterion would 
have included by far less short GRBs in our sample, as their total energies 
are much smaller than those of the long events, while their peak fluxes span 
a broader range. With 18 short GRBs out of 52 selected bright bursts ($\sim$30\%), 
the first year of GBM sample contains many more bright short and hard 
bursts than K06. In fact, only 17 out of 350 BATSE bright bursts are 
short, representing $\sim$5\%\ of the sample. Another reason for the 
difference between the two samples is likely the improved trigger 
algorithms implemented in the GBM FSW. BATSE had three time--interval 
algorithms (based on one energy interval), while GBM currently employs 
28 trigger algorithms at various time and energy channel combinations. 
These algorithms have vastly improved the capability of the instrument 
to trigger on short {\it and hard} GRBs, compared to BATSE. 
\subsection{Correlation among Spectral Parameters -- Empirical Relations}
Empirical correlations among spectral parameters have been
previously found with smaller GRB samples either within individual
bursts or for collections of time--resolved parameters of multiple events.
K06 found no indication of global correlations
among the time--integrated spectral parameters of their BEST sample
(discussed in \S\ \ref{Sec_Comp_BATSE}), while they found strong
correlations among the time--resolved spectral parameters. They also 
noted that it is best to look for parameter correlations within 
{\it individual} bursts to eliminate
possible effects due to cosmological redshift that
varies from burst to burst. Since no time--resolved spectral analysis 
was performed in this work, we limit our correlation analysis to 
comparisons of the low-- and high--energy spectral
parameters against \Epeak\ values and against each other.  

\begin{figure}
\centering
\begin{tabular}{c}
\includegraphics[width=0.48\textwidth,bb=0 0 590 482,clip]{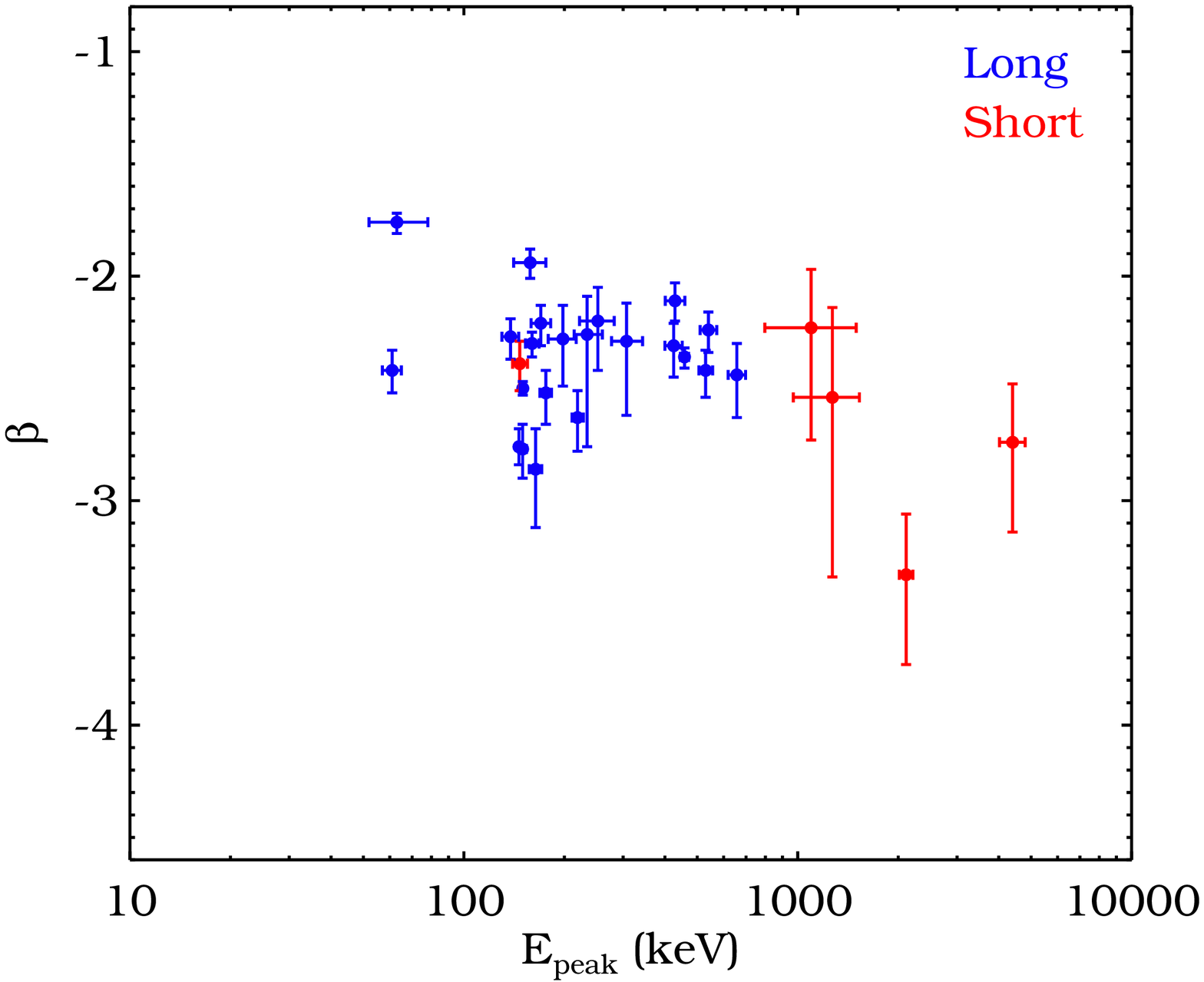} \\
\includegraphics[width=0.48\textwidth,bb=0 0 590 482,clip]{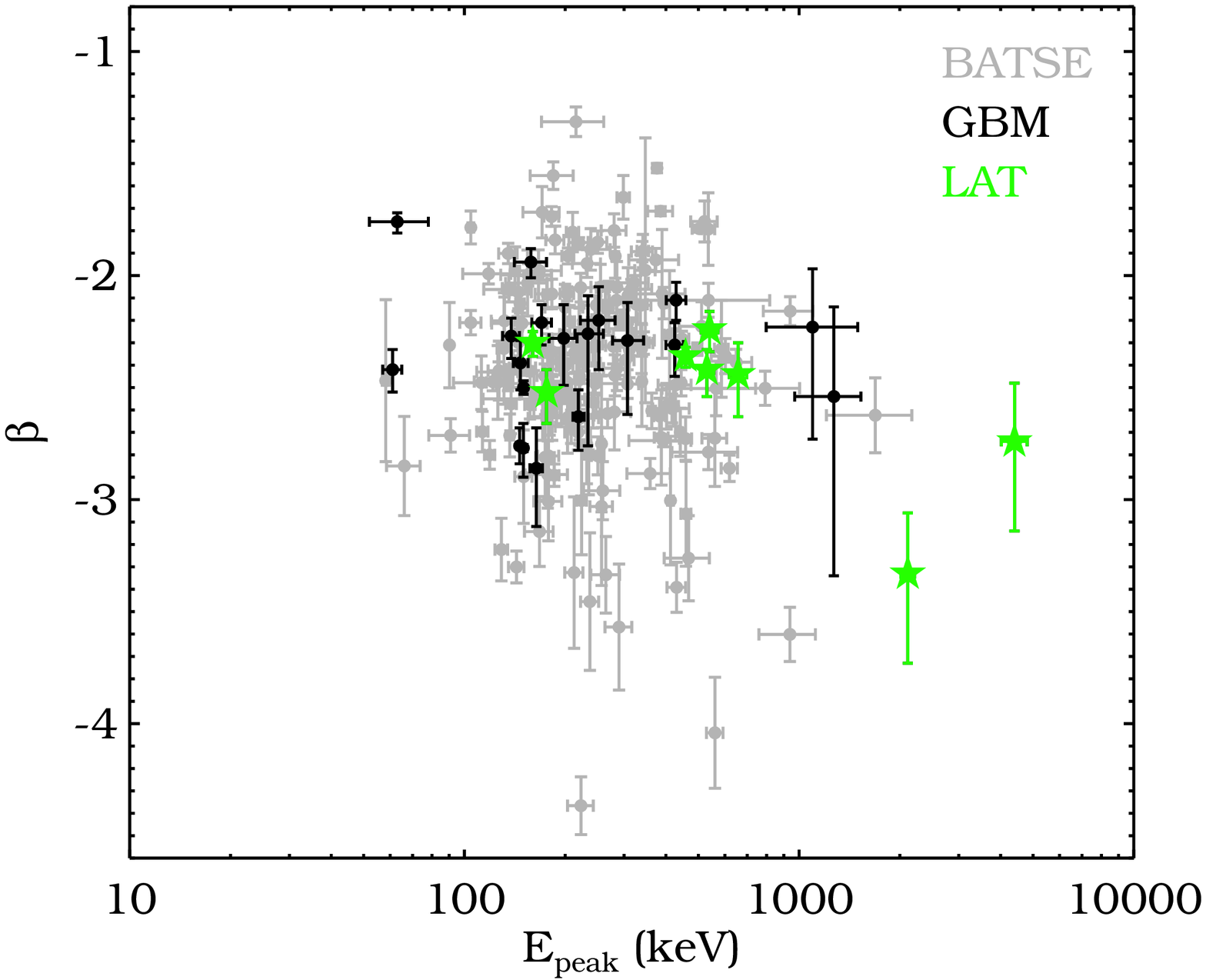} 
\end{tabular}
\caption{Scatter plot of the GBM spectral parameter pair \Epeak--$\beta$
for 22 long GRBs and 5 short GRBs (blue and red data points
in the top panel). Gray data points in the bottom panel represent
350 GRBs from the K06 sample. Bursts fully or marginally detected by the LAT are marked
with green stars.}
\label{Plot_Epeak_Beta}
\end{figure}

Figure~\ref{Plot_Epeak_Beta} shows the scatter plot of the {\it Band} 
function high--energy index $\beta$ {\it versus} \Epeak\ for the GBM 
data alone (top panel), and for the combined data sets of GBM and BATSE 
(bottom panel). The most distinct differences between the two samples is 
the larger \Epeak\ span of the GBM data, and the larger $\beta$ spread 
of the BATSE data.  The reasons for the former have been elaborated in 
\S\ \ref{Sec_Comp_BATSE}; the latter is the effect again of the 
K06 selection for fluence and not for hardness (as in the GBM sample) 
and constitutes, therefore, a more representative characteristic of the 
GRB population as a whole. The GBM subsample of the bright, hard events, 
also includes -- not quite unexpectedly -- the majority of the LAT GRBs, 
and falls well within the BATSE $\beta$ range. 

\begin{figure}
\centering
\begin{tabular}{cc}
\includegraphics[width=0.45\textwidth,bb=0 0 590 482,clip]{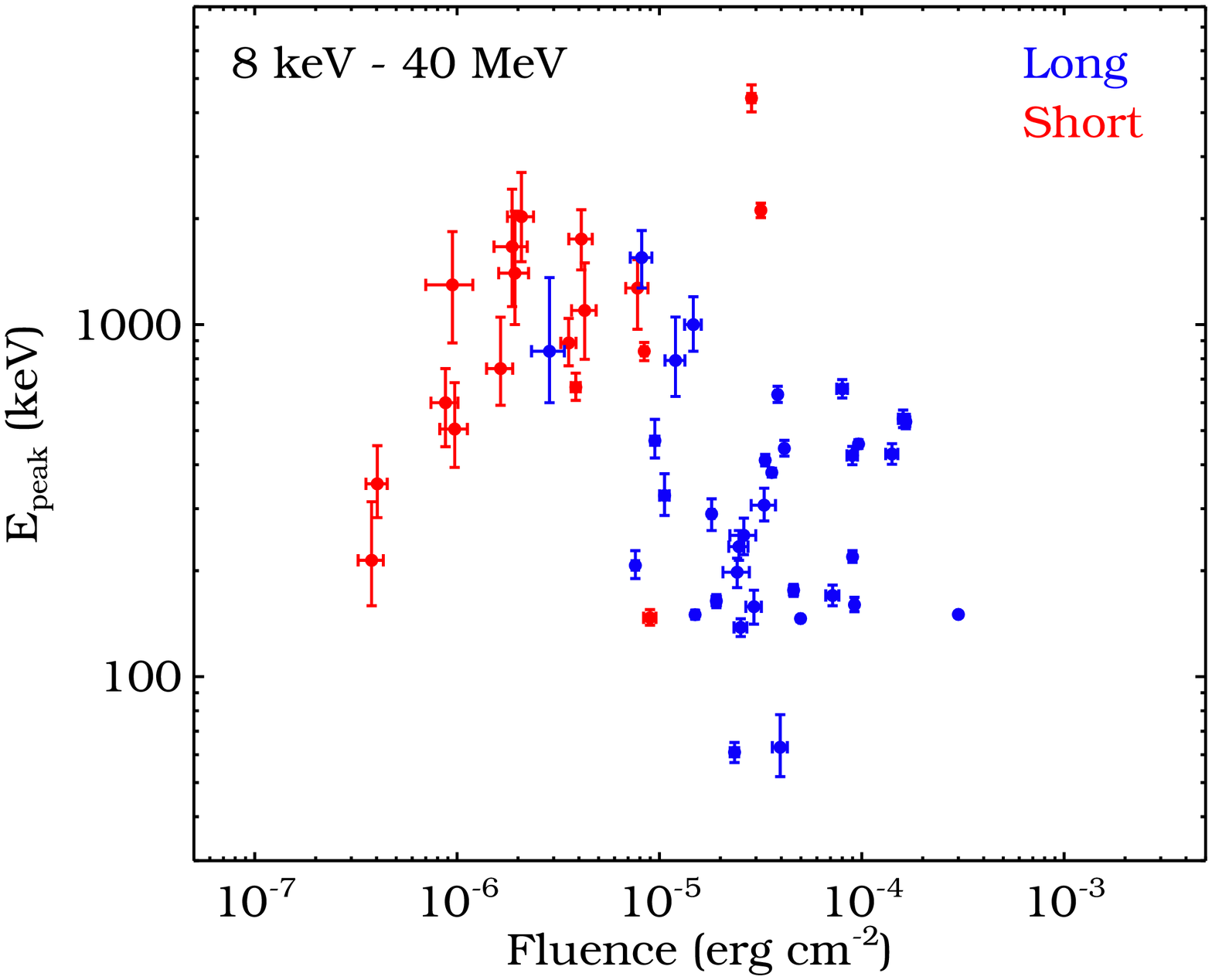} &
\includegraphics[width=0.45\textwidth,bb=0 0 590 482,clip]{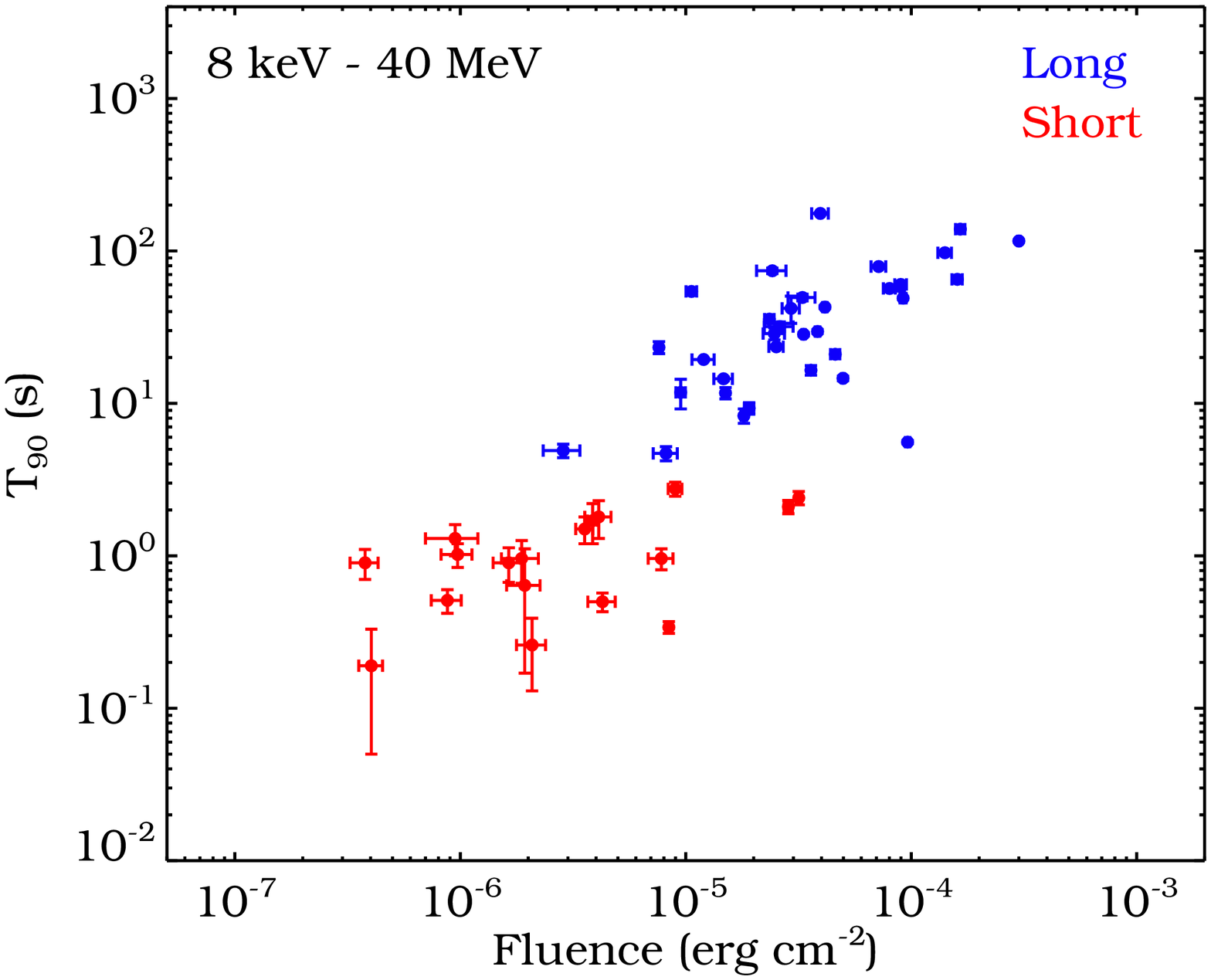} \\
\includegraphics[width=0.45\textwidth,bb=0 0 590 482,clip]{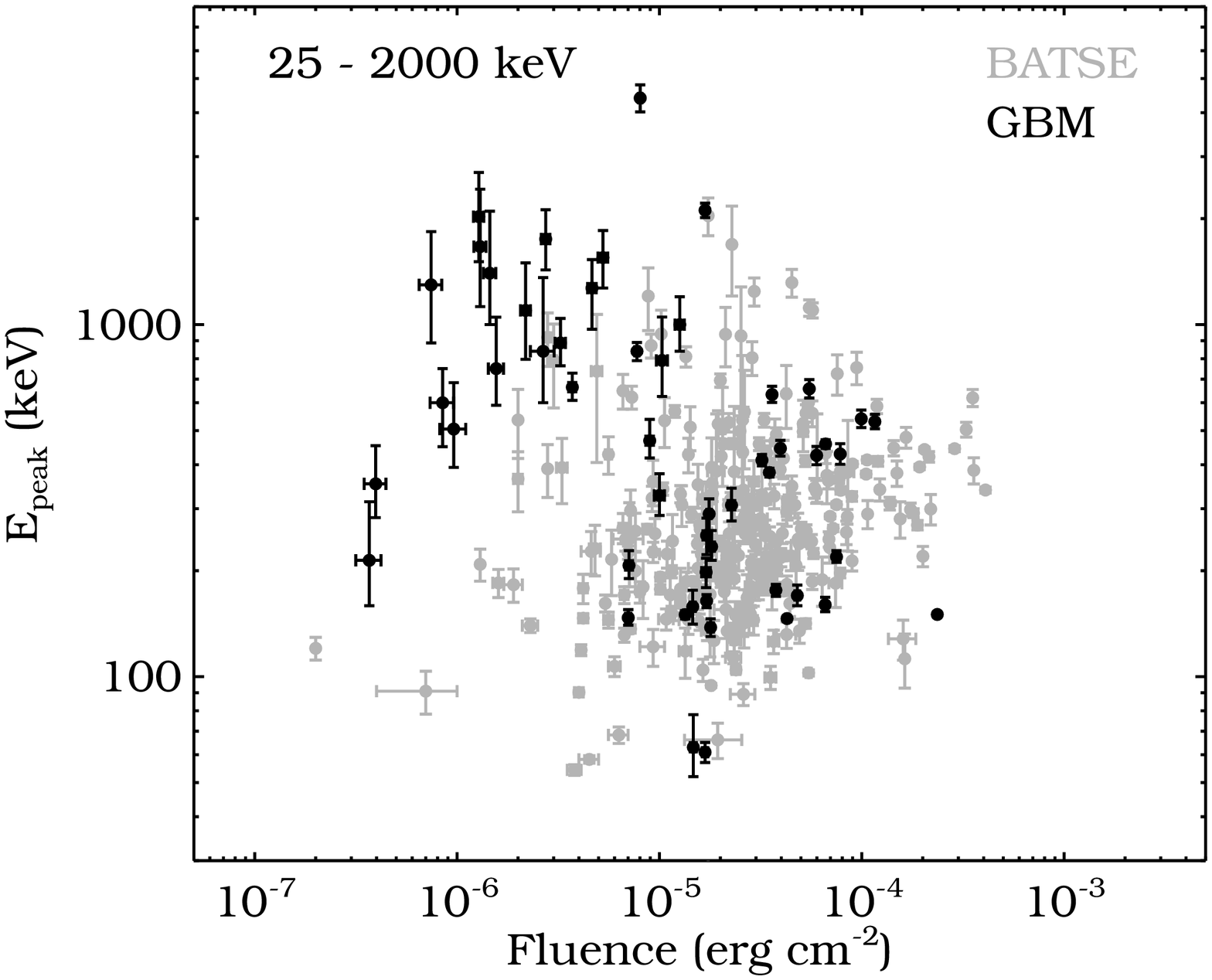} &
\includegraphics[width=0.45\textwidth,bb=0 0 590 482,clip]{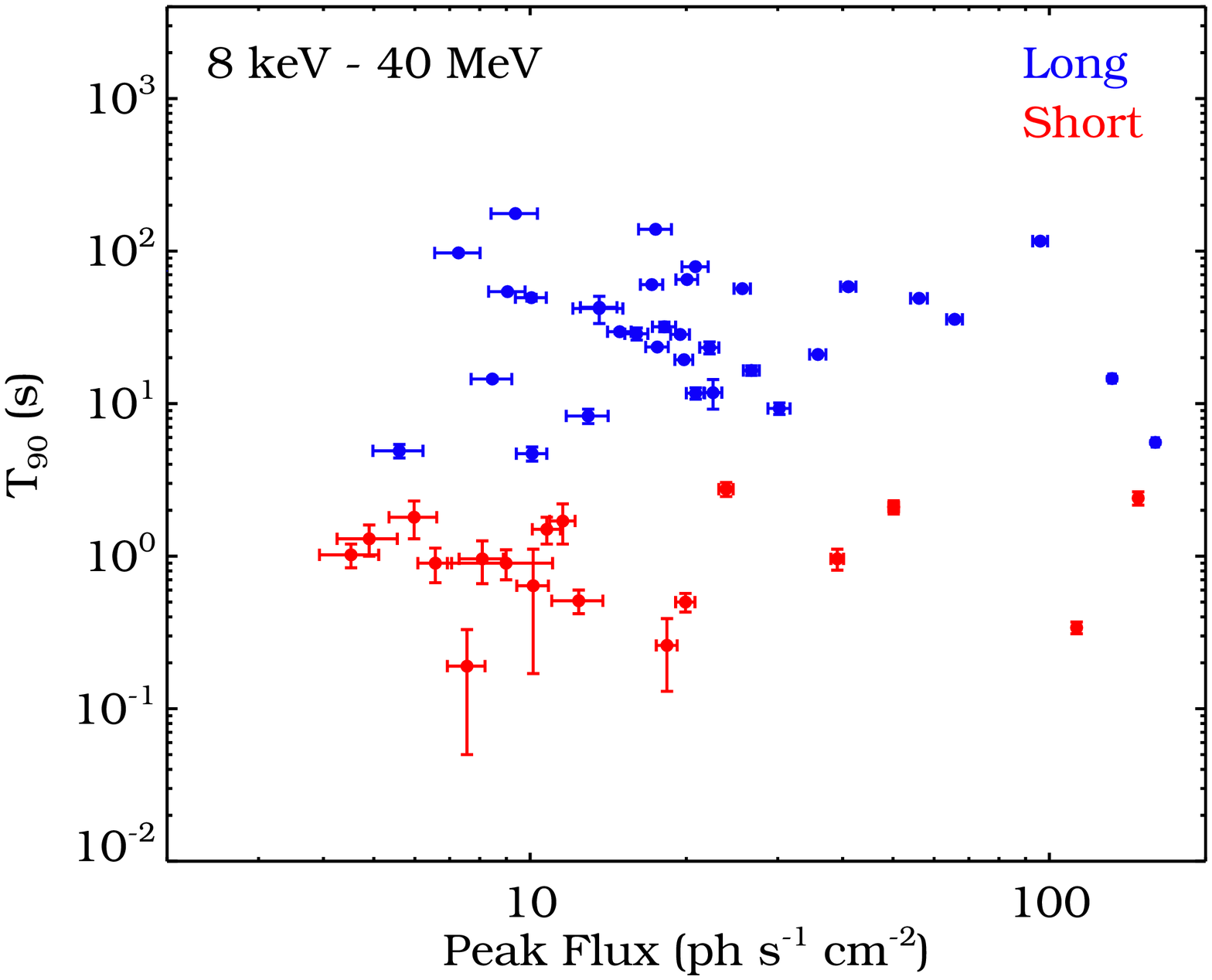}
\end{tabular}
\caption{{\it Left panels}: Distribution of \Epeak\ {\it versus}
energy fluence calculated over the entire GBM energy range (top left panel, 
8\,keV--40\,MeV) and over the BATSE energy range (bottom left panel, 25--2000\,keV). 
Blue and red data points represent long and short GRBs, respectively; 
Grey data points represent 350 GRBs from the K06 sample. 
{\it Right panels}: Distribution of \TN\ (50--300\,keV) {\it versus}
energy fluence (8 keV--40\,MeV, top right panel)
and {\it versus}
128\,ms peak photon fluxes (8\,keV--40\,MeV, bottom right panel).}
\label{Plot_Epeak_Fluence}
\end{figure}

Figure~\ref{Plot_Epeak_Fluence} displays the \Epeak\ and \TN\ distributions against 
energy fluences and peak fluxes measured over the entire GBM energy band, i.e., 
8\,keV--40~MeV. \Epeak\ is plotted against fluences for short and long GBM GRBs 
(red and blue data points, top left panel) and for GBM and BATSE bright GRBs
(black and grey data points, bottom left panel). Short, low--fluence bursts show 
higher \Epeak\ values, while long, high--fluence bursts tend to have lower 
\Epeak\ values. The distribution of \TN\ measured in the standard BATSE 
energy range of 50--300\,keV {\it versus} energy fluence in the GBM  
8\,keV--40\,MeV band is shown in the top right panel of
Figure~\ref{Plot_Epeak_Fluence}. This panel clearly exhibits the 
trend already shown with the \sw\ data that short bursts have lower
fluences than longer ones and that high fluences unambiguously correspond
to longer durations \citep{PIZ09}. The lower right panel exhibits that both long 
and short GRBs have a similar (broad) peak flux range. 

\begin{figure}
\centering
\includegraphics[width=0.5\textwidth,bb=0 0 590 482,clip]{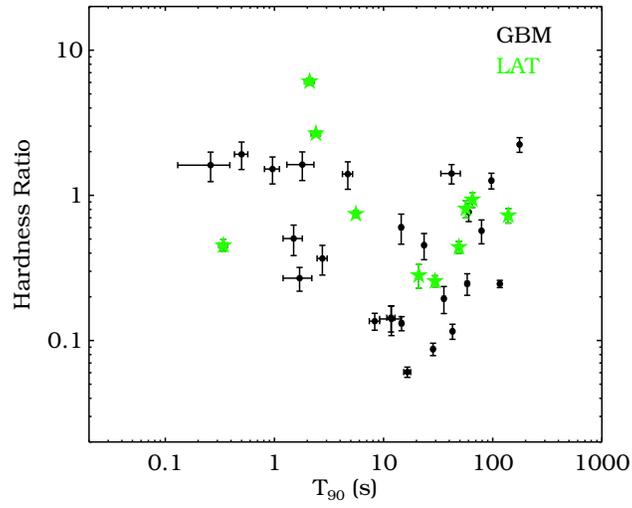}
\caption{Distribution of hardness ratios {\it versus} \TN\ (50--300\,keV). 
The hardness ratios are defined as the ratio of counts collected between 
1--40\,MeV over those collected between 8--1000\,keV. 
Bursts fully or marginally detected by the LAT are marked
with green stars.}
\label{Plot_T90_HR}
\end{figure}

Finally, we plot in Figure~\ref{Plot_T90_HR} the distribution of 
hardness ratios {\it versus} \TN. The hardness ratios are defined 
by the ratio of counts collected in the BGO energy range over 
those collected in the NaI energy range (1000--40000/8--1000, in keV units). 
Although the current sample is not very large, it allows us to distinguish 
that shorter bursts in general tend to have harder spectra than the long ones.  

Amati et al. (2002) reported a relation of \Eiso\ {\it versus} \Epeak\ using 
the BATSE data set. However, \citet{NAK05a} and \citet{BAN05} showed that their 
results may have suffered from strong  selection effects and 
were inconsistent with a larger set of GRB data obtained with BATSE. 
We tested here the Amati relation for those bursts included in 
the BGO--bright burst sample with a redshift measurement. 
We only had 7 cases, namely six long GRBs (GRB~080916C, 
$z=4.35$ \citep{GRE09}; GRB~090102, $z=1.55$; GRB~090323, $z=3.57$; GRB~090328, 
$z=0.74$; GRB~090424, $z=0.54$; and GRB~090618, $z=0.54$) and one 
short burst (GRB~090510, $z=0.90$). We note that long bursts nicely 
follow the \Epeak--\Eiso\ relation, as was also recently pointed out 
by \citet{AMA10}, and the only outlier is the short burst. A recent study 
by Goldstein et al. (2011) explores the Amati relationship in detail 
using the entire GBM data set. 
\section{Conclusions}\label{Sec_Concl}
We have studied here a sample of 52 bright and hard GRBs collected 
during the first year of the GBM operation. We have performed temporal 
and time--integrated spectral analysis of all these events and studied 
the distributions and evolution of the derived parameters. 
The new spectral capabilities afforded with the GBM BGO detectors 
have enabled us to produce a predicting filter using GBM data alone 
of potential GRB detections with the LAT on \fe. This filter will be 
implemented in the GBM FSW and alerts will be distributed to the wide 
scientific community to allow timely multi--wavelength follow up 
observations and, thus, broadband spectral energy distribution studies in GRBs. 

Our temporal evolution analysis has, for the first time, 
extended the duration--energy relationship (found earlier 
in the BATSE data) to the MeV energy range. Although the 
LAT GeV detections seem to be longer in some GRBs
\citep[even up to hundreds of seconds, as in the case of GRB 090323
and GRB 090328, see][]{ABD11b},
there seems to be a single power law relation 
(of index $-0.4$) between duration and energy in the keV to MeV 
prompt $gamma-$ray emission. Whether the GeV emission seen 
with the LAT in several of these GRBs is related to 
the prompt or the afterglow emission is still an open question, which requires 
more data for definite conclusions.

Finally, we show that the novel GBM trigger algorithms have improved 
the collection of short and hard GRBs, compared to the BATSE sample. 
We confirm that their spectral parameter distributions are overall 
similar to those of the K06 sample, and that short GRBs are in general 
harder than longer events. The small subsample of GRBs with known 
distances in our data, is not sufficient to test the various 
empirical relations in the literature. 
%
%
%



\acknowledgments
Support for the German contribution to GBM was provided by the
Bundesministerium f\"ur Bildung und Forschung (BMBF) via the Deutsches Zentrum
f\"ur Luft-- und Raumfahrt (DLR) under contract number 50 QV 0301.
A.v.K.  was supported by the Bundesministeriums f\"ur Wirtschaft und Technologie
(BMWi) through DLR grant 50 OG 1101.
A.J.v.d.H. was supported by NASA grant NNH07ZDA001--GLAST.
S.MB. acknowledges support of the European Union Marie Curie European 
Reintegration Grant within the 7th Program under contract number 
PERG04--GA--2008--239176.

\clearpage

%
\clearpage
%
%
\clearpage
%
%
\clearpage
%
%
\clearpage
%
%
\clearpage
%
%
\clearpage
%
%
\clearpage
%
\renewcommand{\tabcolsep}{10pt}
\renewcommand{\tabcolsep}{5pt}
%
\clearpage
%
\renewcommand{\tabcolsep}{10pt}
\renewcommand{\tabcolsep}{5pt}
%
\clearpage
%
%
\clearpage
%

\clearpage


\renewcommand{\tabcolsep}{4pt}
\renewcommand{\baselinestretch}{0.7}
\begin{deluxetable}{lccccccr@{}lr@{}l}
\tabletypesize{\scriptsize}
\tablecaption{Basic properties of 52 bright GRBs \label{Tab_Prop_All}}
\tablewidth{0pt}
\tablehead{
\noalign{\smallskip}
\colhead{GBM}           & 
\colhead{GRB}           &  
\colhead{Trig.~Time}    & 
\colhead{NaI}           & 
\colhead{BGO}           & 
\colhead{LAT Angle}     & 
\colhead{Data}          &
\multicolumn{4}{c}{Time Interval$^{\rm a}$} \\
\colhead{Trig.~\#}      & 
\colhead{Name}          &  
\colhead{($T_0$, MET)}   & 
\colhead{Det.}          & 
\colhead{Det.}          & 
\colhead{(deg)}         & 
\colhead{Type}             &
\multicolumn{2}{c}{Start} & 
\multicolumn{2}{c}{Stop}  \\    
\noalign{\smallskip}
\colhead{(1)}      & 
\colhead{(2)}      &  
\colhead{(3)}      & 
\colhead{(4)}      & 
\colhead{(5)}      & 
\colhead{(6)}      & 
\colhead{(7)}      &
\multicolumn{2}{c}{(8)} & 
\multicolumn{2}{c}{(9)}   
}
\startdata
\noalign{\smallskip}
080723.557	&	080723B	&	238512142	&	4			&	0	 &	107				&	CSPEC	&	  0&.004	&	60&.161	  \\
080723.985	&	$-$			&	238549063	&	5,2		&	0	 &	113				&	CSPEC	&	 -2&.304	&	50&.945	  \\
080725.541	&	$-$			&	238683564	&	6,7		&	1	 &	50				&	TTE		&	 -0&.064	&	  0&.384	\\
080802.386	&	080802	&	239361311	&	4,5		&	0	 &	125				&	TTE		&	 -0&.064	&	  0&.448	\\
080807.993	&	080807	&	239845833	&	0,1,2	&	0	 &	74				&	CSPEC	&	 -1&.376	&	 21&.152	\\
080816.989	&	080816B	&	240623035	&	b,7		&	1	 &	70				&	TTE		&	 -0&.064	&	  4&.480	\\
080817.161	&	080817A	&	240637931	&	2,5		&	0	 &	80				&	CSPEC	&	  0&.004	&	 60&.417	\\
080825.593	&	080825C	&	241366429	&	9,a		&	1	 &	60				&	CSPEC	&	  0&.004	&	 25&.216	\\
080905.499	&	080905A	&	242308736	&	6,7		&	1	 &	28				&	TTE		&	 -0&.064	&	  1&.024	\\
080906.212	&	080906B	&	242370312	&	0,1,3	&	0	 &	32				&	CSPEC	&	  0&.004	&	  3&.712	\\
080916.009	&	080916C	&	243216766	&	3,4		&	0	 &	52				&	CSPEC	&	  0&.004	&	 70&.145	\\
080925.775	&	080925	&	244060556	&	6,7		&	1	 &	38				&	CSPEC	&	  0&.004	&	 25&.856	\\
081006.604	&	081006	&	244996175	&	0,3		&	0	 &	16				&	TTE		&	 -0&.384	&	  3&.392	\\
081009.690	&	$-$			&	245262818	&	8,b		&	1	 &	96				&	CSPEC	&	 -2&.688	&	 40&.321	\\
081012.045	&	081012B	&	245466323	&	9,a		&	1	 &	66				&	TTE		&	 -0&.128	&	  0&.768	\\
081024.891	&	081024B	&	246576161	&	6,9		&	1	 &	16				&	TTE		&	 -0&.128	&	  0&.128	\\
081101.532	&	081101B	&	247236325	&	5,2		&	0	 &	116				&	CSPEC	&	  0&.003	&	  8&.704	\\
081110.601	&	081110	&	248019944	&	7,8		&	1	 &	67				&	TTE		&	 -0&.192	&	 12&.096	\\
081121.858	&	081121	&	248992528	&	a,b		&	1	 &	140				&	CSPEC	&	  0&.003	&	 21&.504	\\
081122.520	&	081122	&	249049693	&	0,1		&	0	 &	21 (ARR)	&	CSPEC	&	  0&.002	&	 25&.600	\\
081125.496	&	081125	&	249306820	&	a,b		&	1	 &	126				&	CSPEC	&	  0&.003	&	 10&.368	\\
081126.899	&	081126	&	249428050	&	0,1		&	0	 &	18				&	CSPEC	&	-12&.160	&	 40&.065	\\
081129.161	&	081129	&	249623525	&	a,b		&	1	 &	118				&	CSPEC	&	 -2&.944	&	 28&.800	\\
081207.680	&	081207	&	250359527	&	9,a		&	1	 &	56				&	CSPEC	&	  0&.003	&	100&.354	\\
081209.981	&	081209	&	250558317	&	8,b		&	1	 &	107				&	TTE		&	 -0&.056	&	  0&.256	\\
081215.784	&	081215A	&	251059717	&	9,a		&	1	 &	89				&	CSPEC	&	  0&.004	&	  7&.424	\\
081216.531	&	081216	&	251124240	&	8,b		&	1	 &	99				&	TTE		&	 -0&.128	&	  0&.960	\\
081224.887	&	081224	&	251846276	&	6,9		&	1	 &	17 (ARR)	&	CSPEC	&	  0&.002	&	 16&.544	\\
081226.509	&	081226B	&	251986391	&	6,7		&	1	 &	22				&	TTE		&	 -0&.064	&	  0&.192	\\
081231.140	&	081231	&	252386462	&	6,9		&	1	 &	21				&	CSPEC	&	  0&.003	&	 28&.672	\\
090102.122	&	090102	&	252557732	&	a,b		&	1	 &	87				&	CSPEC	&	  0&.003	&	 28&.928	\\
090131.090	&	090131	&	255060563	&	9			&	1	 &	40				&	CSPEC	&	  0&.003	&	 38&.145	\\
090217.206	&	090217	&	256539404	&	6,9,7	&	1	 &	34				&	CSPEC	&	  0&.003	&	 29&.824	\\
090219.074	&	090219	&	256700780	&	5,2		&	0	 &	137				&	TTE		&	 -0&.064	&	  0&.576	\\
090227.310	&	090227	&	257412359	&	0,3		&	0	 &	20				&	CSPEC	&	  0&.003	&	 15&.232	\\
090227.772	&	090227B	&	257452263	&	1,2,5	&	0	 &	72 (ARR)	&	TTE		&	 -0&.128	&	  0&.384	\\
090228.204	&	090228	&	257489602	&	0,1,3	&	0	 &	16				&	TTE		&	 -0&.128	&	  0&.512	\\
090305.052	&	090305B	&	257908477	&	0,3,1	&	0	 &	40				&	TTE		&	 -0&.128	&	  1&.344	\\
090308.734	&	090308B	&	258226586	&	3,7,4	&	0,1	&	50				&	TTE		&	  0&.000	&	  1&.536	\\
090323.002	&	090323	&	259459364	&	9			&	1	  &	53 (ARR)	&	CSPEC	&	  0&.003	&	 71&.681	\\
090328.401	&	090328	&	259925808	&	7,8		&	1	  &	63 (ARR)	&	CSPEC	&	  0&.003	&	 30&.720	\\
090328.713	&	090328B	&	259952826	&	9,10	&	1	  &	74 (ARR)	&	TTE		&	 -0&.064	&	  0&.128	\\
090330.279	&	090330	&	260088144	&	7,9,b	&	1	  &	50				&	CSPEC	&	-38&.913	&	 31&.745	\\
090424.592	&	090424	&	262275130	&	7,8,b	&	1	  &	71				&	CSPEC	&	  0&.002	&	 14&.592	\\
090429.753	&	090429D	&	262721039	&	0,1		&	0	  &	33				&	TTE		&	 -0&.128	&	  0&.512	\\
090510.016	&	090510	&	263607781	&	6,7,9	&	1	  &	13 (ARR)	&	TTE		&	  0&.512	&	  1&.024	\\
090528.516	&	090528B	&	265206153	&	7,8		&	1	  &	65				&	CSPEC	&	  0&.003	&	 80&.897	\\
090531.775	&	090531B	&	265487758	&	6,7,9	&	1	  &	26				&	TTE		&	 -0&.128	&	  0&.832	\\
090618.353	&	090618	&	267006508	&	4			&	0	  &	130				&	CSPEC	&	  8&.704	&	125&.442	\\
090620.400	&	090620	&	267183385	&	6,7,b	&	1	  &	60				&	CSPEC	&	  0&.003	&	 11&.520	\\
090623.107	&	090623	&	267417259	&	7,8,b &	1	  &	73				&	CSPEC	&	 -1&.920	&	 49&.281	\\
090626.189	&	090626	&	267683530	&	0			&	0	  &	15	      &	CSPEC	&	  0&.003	&	 48&.897	\\
\enddata
\tablecomments{
$^{\rm a}$ The time range values are given in $s$ relative to the trigger time $T_0$.
They represent the interval used for the time--integrated spectral analysis}
\end{deluxetable}

\clearpage

\renewcommand{\tabcolsep}{6pt}
\begin{deluxetable}{cccccccccc}
\tabletypesize{\small}
\tablecaption{Duration {\it versus} Energy parameters \label{Tab_Dur_En}}
\tablewidth{0pt}
\tablehead{
\noalign{\smallskip}
\multirow{2}{*}{Energy Range} & \multicolumn{3}{c}{N. of bursts} & & \multicolumn{2}{c}{$\langle{\rm T}_{90} (50-300 \,{\rm keV}) \rangle$} & & \multicolumn{2}{c}{$\langle \alpha_{90} \rangle$} \\
\noalign{\smallskip}
\cline{2-4}\cline{6-7}\cline{9-10}
\noalign{\smallskip}
  & T &  L  & S  & &  L  &  S  & & L  &  S }
\startdata
\noalign{\smallskip}
300 keV\,--\, 2 MeV  &   9  &   6  &   3   &  & 43.1$\,\pm\,$0.7  &  1.08$\,\pm\,$0.18  &  & $-0.41\,\pm\,$0.04 & $-0.42\,\pm\,$0.17  \\
300 keV\,--\, 5 MeV  &   4  &   2  &   2   &  & 52.9$\,\pm\,$0.5  &  0.92$\,\pm\,$0.15  &  & $-0.38\,\pm\,$0.04 & $-0.42\,\pm\,$0.17  \\
300 keV\,--\,10 MeV  &   6  &   4  &   2   &  & 76.7$\,\pm\,$0.6  &  2.24$\,\pm\,$0.16  &  & $-0.32\,\pm\,$0.02 & $-0.40\,\pm\,$0.12  \\
\enddata
\end{deluxetable}
\renewcommand{\tabcolsep}{5pt}

\clearpage


\renewcommand{\tabcolsep}{4pt}
\renewcommand{\baselinestretch}{0.7}
\begin{deluxetable}{ccccccccc}
\rotate	
\tabletypesize{\small}
\tablecaption{Summary of time--integrated spectral fit results OF 52 bright GRBs \label{Tab_Spec_MAIN}}
\tablewidth{0pt}
\tablehead{
\noalign{\smallskip}
\colhead{GBM}             & 
\colhead{Best}            & 
\colhead{$A$}             & 
\colhead{\Epeak}          & 
\multirow{2}{*}{index}    & 
\multirow{2}{*}{$\alpha$} & 
\multirow{2}{*}{$\beta$}  & 
\colhead{A${\rm{_{eff}}}$}           & 
\colhead{CSTAT/}           \\ 
\colhead{Trig.~\#}   & 
\colhead{Model}      & 
\colhead{$({\rm ph.\,s^{-1}\,cm^{-2}\,keV^{-1}})$}   & 
\colhead{(keV)}      & 
\colhead{ }          & 
\colhead{ }          & 
\colhead{ }          & 
\colhead{Corr.}      & 
\colhead{DOF}        \\ 
\colhead{(1)}          & 
\colhead{(2)}          & 
\colhead{(3)}          & 
\colhead{(4)}          & 
\colhead{(5)}          & 
\colhead{(6)}          & 
\colhead{(7)}          & 
\colhead{(8)}          & 
\colhead{(9)}            
}
\startdata
\noalign{\smallskip}
080723.557	&	Band	&	0.035		$\pm$	0.0008	&	 219	$\left(_{-8}^{+9}\right)$  &		$\ldots$  &	$-$0.858	$\left(_{-0.023	}^{+	0.023}\right)$  &	-2.63	$\left(_{-0.15}^{+0.12}\right)$  &	0.83	&	384/235	\\
080723.985	&	Comp	&	0.0111	$\pm$	0.0003	&	 445	$\left(_{-22}^{+24}\right)$    &	-0.949	$\left(_{-0.026}^{+0.027}\right)$  &	$\ldots$  &		$\ldots$  &	$\ldots$	&	496/359	\\
080725.541	&	Comp	&	0.0135	$\pm$	0.0009	&	1670	$\left(_{-540}^{+760}\right)$  &	-0.89	$\left(_{-0.11}^{+0.14}\right)$  &		$\ldots$  &		$\ldots$  &	$\ldots$	&	362/359	\\
080802.386	&	Comp	&	0.0166	$\pm$	0.0025	&	 600	$\left(_{-150}^{+150}\right)$  &	-0.65	$\left(_{-0.18}^{+0.21}\right)$  &		$\ldots$  &		$\ldots$  &	$\ldots$	&	402/358	\\
080807.993	&	Comp	&	0.0043	$\pm$	0.0002	&	 790	$\left(_{-170}^{+260}\right)$  &	-1.01	$\left(_{-0.06}^{+0.07}\right)$  &		$\ldots$  &		$\ldots$  &	$\ldots$	&	577/481	\\
080816.989	&	Comp	&	0.0039	$\pm$	0.0002	&	1550	$\left(_{-280}^{+300}\right)$  &	-0.51	$\left(_{-0.11}^{+0.12}\right)$  &		$\ldots$  &		$\ldots$  &	$\ldots$	&	390/360	\\
080817.161	&	Band	&	0.0147	$\pm$	0.0004	&	 425	$\left(_{-25}^{+26}\right)$  	&		$\ldots$  &	-0.99	$\pm$ 0.02  &	-2.31	$\left(_{-0.14}^{+0.10}\right)$  &	$\ldots$	&	544/361	\\
080825.593	&	Band	&	0.0617	$\pm$	0.0033	&	 176	$\pm$ 7  											&		$\ldots$  &	-0.64 $\pm$ 0.04	  &	-2.52	$\left(_{-0.14}^{+0.10}\right)$  &	0.80	&	418/357	\\
080905.499	&	Comp	&	0.0090	$\pm$	0.0016	&	 500	$\left(_{-110}^{+180}\right)$  &	-0.20	$\left(_{-0.29}^{+0.4}\right)$  &		$\ldots$  &		$\ldots$  &	$\ldots$	&	471/361	\\
080906.212	&	Band	&	0.1152	$\pm$	0.0110	&	 147	$\left(_{-7}^{+8}\right)$  		&		$\ldots$  &	-0.37	$\pm$ 0.06   &	-2.39	$\left(_{-0.12}^{+0.10}\right)$  &	$\ldots$	&	522/479	\\
080916.009	&	Band	&	0.0166	$\pm$	0.0002	&	 540	$\left(_{-30}^{+32}\right)$  	&		$\ldots$  &	-1.06	$\pm$ 0.02  &	-2.24	$\left(_{-0.10}^{+0.08}\right)$  &	$\ldots$	&	533/358	\\
080925.775	&	Band	&	0.0272	$\pm$	0.0016	&	138	$\pm$ 8  												&		$\ldots$  &	-0.96	$\pm$ 0.04  &	-2.27	$\left(_{-0.10}^{+0.08}\right)$  &	$\ldots$	&	430/359	\\
081006.604	&	Comp	&	0.0038	$\pm$	0.0004	&	840	$\left(_{-240}^{+520}\right)$  &	-0.43	$\left(_{-0.26}^{+0.3}\right)$  &		$\ldots$  &		$\ldots$  &	$\ldots$	&	423/362	\\
081009.690	&	Band	&	0.0296	$\pm$	0.0143	&	63	$\left(_{-11}^{+15}\right)$  		&		$\ldots$  &	-0.53	$\left(_{-0.24}^{+0.3}\right)$  &	-1.76	$\left(_{-0.05}^{+0.04}\right)$  &	0.70	&	502/358	\\
081012.045	&	Comp	&	0.0123	$\pm$	0.0011	&	750	$\left(_{-160}^{+300}\right)$  	&	-0.44	$\left(_{-0.20}^{+0.23}\right)$  &		$\ldots$  &		$\ldots$  &	 $\ldots$	&	368/355	\\
081024.891	&	Comp	&	0.0096	$\pm$	0.0014	&	1300	$\left(_{-410}^{+540}\right)$  &	-0.46	$\left(_{-0.22}^{+0.3}\right)$  &		$\ldots$  &		$\ldots$  &	$\ldots$	&	326/359	\\
081101.532	&	Comp	&	0.0251	$\pm$	0.0008	&	290	$\pm$ 30  											&	-0.686	$\pm$ 0.04  										&		$\ldots$  &		$\ldots$  &	$\ldots$	&	360/360	\\
081110.601	&	Comp	&	0.0105	$\pm$	0.0004	&	470	$\left(_{-50}^{+70}\right)$  &	-1.064	$\left(_{-0.043}^{+0.046}\right)$  &		$\ldots$  &		$\ldots$  &	$\ldots$	&	410/359	\\
081121.858	&	Band	&	0.0266	$\pm$	0.0049	&	158	$\left(_{-17}^{+18}\right)$  &		$\ldots$  &	-0.47	$\left(_{-0.12}^{+0.14}\right)$  &	-1.94	$\left(_{-0.07}^{+0.06}\right)$  &	$\ldots$	&	456/356	\\
081122.520	&	Comp	&	0.0093	$\pm$	0.0006	&	207	$\left(_{-17}^{+21}\right)$  &	-0.90	$\pm$ 0.07  												&		$\ldots$  &		$\ldots$  &	$\ldots$	&	435/360	\\
081125.496	&	Band	&	0.0932	$\pm$	0.0051	&	164	$\pm$ 7  											&		$\ldots$  			&	-0.48	$\pm$ 0.05  		&	-2.86	$\left(_{-0.26}^{+0.18}\right)$  &	0.76	&	408/358	\\
081126.899	&	Comp	&	0.0040	$\pm$	0.0004	&	330	$\left(_{-40}^{+50}\right)$  &	-1.00	$\pm$ 0.07 													&		$\ldots$  &		$\ldots$  &	$\ldots$	&	488/360	\\
\hline
\tablebreak
081129.161	&	Band	&	0.0114	$\pm$	0.0007	&	250	$\pm$ 30  										&		$\ldots$  			&	-0.96	$\left(_{-0.05}^{+0.06}\right)$  &	-2.20	$\left(_{-0.22}^{+0.15}\right)$  &	$\ldots$	&	454/356	\\
081207.680	&	Band	&	0.0095	$\pm$	0.0002	&	430	$\left(_{-28}^{+30}\right)$  &		$\ldots$  &	-0.67	$\pm$ 0.03  &	-2.11	$\left(_{-0.09}^{+0.08}\right)$  &	$\ldots$	&	501/356	\\
081209.981	&	Band	&	0.0347	$\pm$	0.0021	&	1100	$\left(_{-300}^{+400}\right)$  &		$\ldots$  &	-0.68	$\left(_{-0.11}^{+0.14}\right)$  &	-2.23	$\left(_{-0.5}^{+0.26}\right)$  &	$\ldots$	&	349/357	\\
081215.784	&	Band	&	0.1074	$\pm$	0.0014	&	458	$\pm$ 13  											&		$\ldots$  &	-0.71	$\left(_{-0.02}^{+0.02}\right)$  &	-2.36	$\left(_{-0.05}^{+0.04}\right)$  &	$\ldots$	&	519/356	\\
081216.531	&	Band	&	0.0200	$\pm$	0.0009	&	1270	$\left(_{-300}^{+260}\right)$  &		$\ldots$  &	-0.81	$\left(_{-0.05}^{+0.07}\right)$  &	-2.54	$\left(_{-0.80}^{+0.40}\right)$  &	$\ldots$	&	453/357	\\
081224.887	&	Comp	&	0.0378	$\pm$	0.0007	&	380	$\pm$ 11  											&	-0.73	$\pm$ 0.02  												&		$\ldots$  &		$\ldots$  &	0.87	&	442/358	\\
081226.509	&	Comp	&	0.0311	$\pm$	0.0044	&	350	$\left(_{-70}^{+100}\right)$  	&	-0.41	$\left(_{-0.20}^{+0.24}\right)$  &		$\ldots$  &		$\ldots$  &	$\ldots$	&	414/360	\\
081231.140	&	Band	&	0.0149	$\pm$	0.0006	&	234	$\left(_{-20}^{+26}\right)$  &		$\ldots$  &	-1.06	$\pm$ 0.04  					&	-2.26	$\left(_{-0.50}^{+0.17}\right)$  &	$\ldots$	&	472/358	\\
090102.122	&	Comp	&	0.0180	$\pm$	0.0003	&	412	$\left(_{-15}^{+16}\right)$  		&	-0.86	$\pm$ 0.02 												&		$\ldots$  &		$\ldots$  &	0.87	&	442/357	\\
090131.090	&	Band	&	0.0321	$\pm$	0.0030	&	61	$\pm$ 4 												&		$\ldots$  &	-1.21	$\left(_{-0.06}^{+0.08}\right)$  &	-2.42	$\left(_{-0.10}^{+0.09}\right)$  &	$\ldots$	&	348/236	\\
090217.206	&	Comp	&	0.0125	$\pm$	0.0002	&	633	$\left(_{-32}^{+35}\right)$  			&	-0.91	$\pm$ 0.02  								&		$\ldots$  &		$\ldots$  &	$\ldots$	&	659/479	\\
090219.074	&	Comp	&	0.0289	$\pm$	0.0115	&	214	$\left(_{-55}^{+100}\right)$  		&	-0.2	$\left(_{-0.5}^{+0.8}\right)$  &		$\ldots$  &		$\ldots$  &	$\ldots$	&	353/361	\\
090227.310	&	Comp	&	0.0055	$\pm$	0.0002	&	1000	$\left(_{-160}^{+200}\right)$  		&	-0.86	$\pm$ 0.06  								&		$\ldots$  &		$\ldots$  &	$\ldots$	&	383/360	\\
090227.772	&	Band	&	0.0762	$\pm$	0.0016	&	2100	$\pm$ 100  												&		$\ldots$  &	-0.51	$\left(_{-0.02}^{+0.03}\right)$  &	-3.33	$\left(_{-0.40}^{+0.27}\right)$  &	$\ldots$	&	548/479	\\
090228.204	&	Comp	&	0.0755	$\pm$	0.0016	&	840	$\pm$ 50  													&	-0.60	$\pm$ 0.03  								&		$\ldots$  &		$\ldots$  &	$\ldots$	&	540/480	\\
090305.052	&	Comp	&	0.0126	$\pm$	0.0006	&	890 $\left(_{-120}^{+150}\right)$  			&	-0.58	$\left(_{-0.09}^{+0.11}\right)$  &		$\ldots$  &		$\ldots$  &	$\ldots$	&	570/480	\\
090308.734	&	Comp	&	0.0193	$\pm$	0.0080	&	664	$\left(_{-50}^{+60}\right)$  			&	-0.53	$\left(_{-0.07}^{+0.08}\right)$  &		$\ldots$  &		$\ldots$  &	$\ldots$	&	688/600	\\
090323.002	&	Band	&	0.0178	$\pm$	0.0003	&	530	$\left(_{-24}^{+26}\right)$  &		$\ldots$  &	-0.81	$\pm$ 0.02 					&	-2.42	$\left(_{-0.12}^{+0.09}\right)$  &	$\ldots$	&	568/237	\\
090328.401	&	Band	&	0.0173	$\pm$	0.0003	&	660	$\pm$ 40 &		$\ldots$  						&	-0.93	$\pm$ 0.02  						&	-2.44	$\left(_{-0.19}^{+0.14}\right)$  &	$\ldots$	&	534/360	\\
090328.713	&	Comp	&	0.0319	$\pm$	0.0016	&	2000	$\left(_{-520}^{+680}\right)$  			&	-0.96	$\pm$ 0.07  							&		$\ldots$  &		$\ldots$  &	$\ldots$	&	376/356	\\
090330.279	&	Band	&	0.0068	$\pm$	0.0005	&	198	$\pm$ 19  &		$\ldots$  &	-0.92	$\left(_{-0.06}^{+0.07}\right)$  &	-2.28	$\left(_{-0.21}^{+0.15}\right)$  & $\ldots$		&	747/477	\\
090424.592	&	Band	&	0.1419	$\pm$	0.0026	&	146.2	$\left(_{-2.9}^{+2.8}\right)$  			&		$\ldots$  &	-0.86	$\pm$ 0.02  					&	-2.76	$\pm$ 0.08  &	0.78	&	843/478	\\
090429.753	&	Comp	&	0.0150	$\pm$	0.0008	&	1400	$\left(_{-400}^{+700}\right)$  &	-1.06	$\left(_{-0.08}^{+0.09}\right)$  &		$\ldots$  &		$\ldots$  &	$\ldots$	&	372/360	\\
\hline
\tablebreak
090510.016	&	Band	&	0.0427	$\pm$	0.0009	&	4400	$\left(_{-380}^{+400}\right)$  &		$\ldots$  &	-0.79	$\pm$ 0.03  &	-2.74	$\left(_{-0.40}^{+0.26}\right)$  &	$\ldots$	&	486/479	\\
090528.516	&	Band	&	0.0170	$\pm$	0.0070	&	170	$\left(_{-11}^{+12}\right)$  &		$\ldots$  &	-1.10	$\pm$ 0.04 			&	-2.21	$\left(_{-0.10}^{+0.08}\right)$  &	$\ldots$	&	553/358	\\
090531.775	&	Comp	&	0.0094	$\pm$	0.0006	&	1750	$\left(_{-320}^{+370}\right)$  &	-0.63	$\left(_{-0.09}^{+0.11}\right)$  &		$\ldots$  &		$\ldots$  &	$\ldots$	&	632/481	\\
090618.353	&	Band	&	0.0717	$\pm$	0.0011	&	150.2	$\pm$ 2.7  											&		$\ldots$  &	-1.12	$\pm$ 0.01 								&	-2.50	$\pm$ 0.03  &	0.78	&	532/233	\\
090620.400	&	Band	&	0.0920	$\pm$	0.0045	&	150	$\pm$ 4  													&		$\ldots$  	&	-0.174	$\pm$ 0.05  &	-2.77	$\left(_{-0.13}^{+0.11}\right)$  &	0.70	&	580/479	\\
090623.107	&	Band	&	0.0084	$\pm$	0.0004	&	307	$\left(_{-30}^{+36}\right)$  &		$\ldots$  &	-0.63	$\left(_{-0.06}^{+0.07}\right)$  &	-2.29	$\left(_{-0.3}^{+0.17}\right)$  &	0.72	&	607/478	\\
090626.189	&	Band	&	0.0440	$\pm$	0.0014	&	160	$\left(_{-7}^{+8}\right)$  &		$\ldots$  &	-1.04	$\pm$ 0.03  				&	-2.30	$\left(_{-0.06}^{+0.05}\right)$  &	0.74	&	353/240
\enddata
\tablecomments{
$^{\rm a}$ The time range values are given in $s$ relative to the trigger time $T_0$.
They represent the interval used for the time--integrated spectral analysis}
\end{deluxetable}

\end{document}